\documentclass[manuscript, nonacm]{acmart}

\usepackage[utf8]{inputenc}

\title[Generation Probabilities Are Not Enough]{Generation Probabilities Are Not Enough: \\Uncertainty Highlighting in AI Code Completions}

\author{Helena Vasconcelos}\thanks{Conducted while HV was an intern at Microsoft Research. GB, AF, QVL, and JWV contributed equally. This is a preprint of an article to appear in TOCHI. DOI: 10.1145/3702320}
\email{helenav@cs.stanford.edu}
\affiliation{%
  \institution{Stanford University}
  \country{USA}
}

\author{Gagan Bansal*}
\affiliation{%
  \institution{Microsoft Research}
  \country{USA}
}

\author{Adam Fourney*}
\affiliation{%
  \institution{Microsoft Research}
  \country{USA}
}

\author{Q. Vera Liao*}
\affiliation{%
  \institution{Microsoft Research}
  \country{USA}
}

\author{Jennifer~Wortman~Vaughan*}
\affiliation{%
  \institution{Microsoft Research}
  \country{USA}
}


\renewcommand{\shortauthors}{Vasconcelos et al.}

\usepackage{amsmath}
\usepackage[utf8]{inputenc} 
\usepackage[T1]{fontenc}    
\usepackage{hyperref}       
\usepackage{url}            
\usepackage{booktabs}       
\usepackage{amsfonts}       
\usepackage{nicefrac}       
\usepackage{microtype}      
\usepackage{xcolor}         
\usepackage{graphicx}
\usepackage{subfigure}
\usepackage{verbatim}
\usepackage{color-edits}

\addauthor{gb}{cyan}

\begin{document}


\begin{abstract}
  Large-scale generative models have enabled the development of AI-powered code completion tools to assist programmers in writing code.
  Like all AI-powered tools, these code completion tools are not always accurate and can introduce bugs or even security vulnerabilities into code if not properly detected and corrected by a human programmer. One technique that has been proposed and implemented to help programmers locate potential errors is to highlight uncertain tokens. However, little is known about the effectiveness of this technique. Through a mixed-methods study with 30 programmers, we compare three conditions: providing the AI system’s code completion alone, highlighting tokens with the lowest likelihood of being generated by the underlying generative model, and highlighting tokens with the highest predicted likelihood of being edited by a programmer. We find that highlighting tokens with the highest predicted likelihood of being edited leads to faster task completion and more targeted edits, and is subjectively preferred by study participants. In contrast, highlighting tokens according to their probability of being generated does not provide any benefit over the baseline with no highlighting. We further explore the design space of how to convey uncertainty in AI-powered code completion tools and find that programmers prefer highlights that are granular, informative, interpretable, and not overwhelming. This work contributes to building an understanding of what uncertainty means for generative models and how to convey it effectively.
\end{abstract}

\maketitle
\section{Introduction}
Large generative models, such GPT-3~\cite{brown2020language},  have made it possible to create previously unimaginable tools for human-AI collaboration for tasks such as writing~\cite{chakrabarty-arxiv2022,clark-iui2018}, generating art or music~\cite{louie-chi2020,liu-uist2022}, and, in particular, computer programming. AI-powered code completion tools such as GitHub's Copilot \cite{copilot_2022}, Amazon's CodeWhisperer \cite{codewhisperer_2022}, and DeepMind's AlphaCode \cite{alphacode_2022}, recommend code completions within integrated development environments (IDEs) to help programmers author software, and are projected to have a profound impact on developer productivity and experience~\cite{copilot-gh2022}. However, these models are imperfect, and are often criticized for their inability to distinguish correct outputs from outputs that are incorrect yet seemingly coherent and plausible~\cite{hallucinating2022,bender2021stochastic}. In the context of code completion, erroneous recommendations can introduce bugs or even security vulnerabilities into a code base~\cite{pearce_2022}. 

For effective human-AI collaboration on code, programmers should be able to detect and correct errors introduced by the code completion tool. This can be challenging as even experts may be susceptible to automation bias, overreliance, and automation-induced complacency \cite{parasuraman2010complacency, wickens2015complacency}. To help operators detect and override errors in medical \cite{cai2019hello,lundberg2018explainable}, legal \cite{angwin2016machine,hayashi2017can}, hiring \cite{liem_2018}, and other high-stakes domains, conveying AI uncertainty and providing explanations are of paramount importance~\cite{bhatt2021uncertainty,bansal-chi21}. However, such decision-support scenarios are typically limited to categorical output, such as a medical diagnosis, and it is not clear how to translate these strategies or findings to generative scenarios with complex natural language outputs. 

Prior works have proposed techniques that help programmers locate potential errors by highlighting uncertain tokens~\cite{ibm, playground}. Similar to in-line spell-check in text editors, highlighted tokens are meant to draw attention to regions of the code that would benefit most from human oversight. However, there are multiple ways of quantifying uncertainty and it is unclear which is most effective.  First, code generation models have a built-in notion of uncertainty---specifically, the likelihood that the model would generate a specific token given its surrounding context~\cite{bengio2003neural,bender2021stochastic}. We refer to this as the ``generation probability,'' and it is this quantity that previous proposals and implementations were based on~\cite{ibm,playground}. However, to the best of our knowledge, there have been no empirical studies exploring the effectiveness of this technique. In this work, we explore another notion of uncertainty directly inspired by programmers' goals: assessing the likelihood that one would need to modify or delete a token in order to solve the given problem. This cannot be obtained directly from the code generation model, but can potentially be approximated through a separate ``edit model'' based on programmers' actions in similar contexts.

With these two notions of uncertainty in mind, we explore whether, and which way of, conveying uncertainty can enable programmers to more quickly and accurately produce code in collaboration with an AI-powered code completion tool.  Through a preregistered mixed-methods study with $N=30$ programmers, we compare three conditions: providing the AI system’s code completion alone, highlighting tokens with the lowest generation probability, and highlighting tokens with the highest predicted likelihood of being edited by a programmer. To implement the final condition, we build a simple version of an edit model to directly predict this likelihood, trained on a closed-world set of programming tasks. While the model we train is simplistic, there are paths to generalize the approach by learning from existing large-scale telemetry data, as discussed in Section~\ref{sec:discussion}.

We find that uncertainty highlighting leads to improved speed, but this outcome depends on which notion of uncertainty is used. Specifically, highlighting tokens with the edit model leads to faster task completion and more targeted edits. Participants are more likely to edit tokens highlighted by the edit model, and also report a higher subjective utility of highlights generated by the edit model compared with those generated using generation probabilities. In contrast, across a variety of measures, highlighting tokens with the lowest generation probability does not provide any benefit over the baseline condition in which no highlights are shown.  

We further explore the design space of uncertainty highlighting for code generation by showing participants alternative design probes. We find that participants prefer highlights that are sufficiently granular to allow them to narrow in on potential errors, but without overwhelming them with unnecessary details---for example, shading is strongly preferred to presenting precise probabilities that may be distracting and slow programmers down. Participants also request explanations for uncertainty highlighting (to help them diagnose the underlying bugs) and control over settings such as the uncertainty highlighting threshold.

\begin{figure}[tb]
    \centering
    \includegraphics[width=\textwidth]{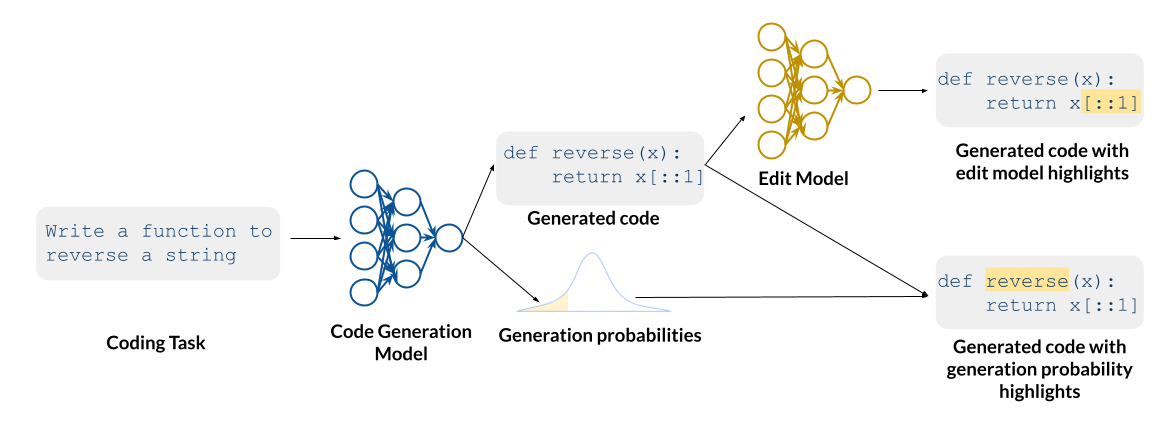}
    \caption{Overview of two approaches to uncertainty highlighting explored in our study: directly predicting the likelihood of an edit (top path), and using generation probabilities (bottom path). In this example, edit model-based highlighting points the programmer to the location of a necessary change (the list comprehension which erroneously has ``1'' in place of ``-1'').  The generation probability-based highlighting points the programmer to the function name, which, while being high variance, does not necessarily have any errors. 
    }
    \label{fig:MainComparisonFigure}
\end{figure}

\section{Related Work} 
AI systems can make mistakes for various reasons, and users may not always be able to detect their errors. This can lead to \emph{overreliance} on AI, which prior work has shown can decrease task performance~\cite{bansal-chi21,jacobs-psych2021},
reduce trust~\cite{yin-chi2019,bansal-chi21},
and worsen user experience~\cite{kocielnik-chi2019}. It is therefore important to help users detect errors in AI predictions, especially in high-stakes domains, such as healthcare and criminal justice, where erroneous predictions can have serious consequences~\cite{green-chi2020}. Previous research has shown that techniques such as communicating uncertainty~\cite{zhang-faact2020}, explaining AI predictions~\cite{gonzalez-acl2021}, and providing appropriate onboarding and user training~\cite{cai-chi2021,lai-chi2020,mozannar-aaai2021} can help users detect errors in AI predictions. Communicating uncertainty, which is our focus, has been shown to be an effective method for improving task performance~\cite{bansal-chi21}, detecting errors~\cite{bansal-chi21,gonzalez-acl2021}, and increasing trust~\cite{bhatt2021uncertainty,zhang-faact2020}. However, it is still an open question how to best present uncertainty, how to compute it, and whether it helps for the increasingly important generative scenario of AI-powered code completion. And, although there might be some insights from prior work on human-human programming, there are unique challenges that human-AI programming raises~\cite{sarkar2022like}. 

\subsection{Computing and Communicating Uncertainty}
There are various techniques for computing and communicating {\em uncertainty} in AI predictions, such as using gauges, percentages, highlights, discretization, and deferring to the user (e.g., withholding uncertain predictions)~\cite{wang-elsevier2021,dhami-elsevier2022}. Post hoc calibration methods and auxiliary models can also be used to compute and improve uncertainty scores~\cite{guo-pmlr2017,niculescu-icml2005,jiang-tacl2021}. However, it is unclear how these techniques from classification-like tasks can be applied to generative tasks or the specific scenario of AI-powered code completion. This lack of generalization may be due to differences in the nature of the tasks. In classification, predictions are typically atomic, structured, and simple, whereas for text or code generation, predictions are often tokens of varying length and are unstructured. 

Even with these difficulties, a growing body of work has emerged that studies and proposes methods for the computation of uncertainty and its effects on the outputs of generative models~\cite{johnson2023ru, khakhar2023pac, kuhn2023semantic, lin2022teaching, kadavath2022language}. These methods and evaluations range from prompting models to output their uncertainty in natural language~\cite{lin2022teaching, liu2023prudent, kotelanski2023methods, singh2023confidencecompetence}, to training models to predict the probability they know the output in addition to the regular objective~\cite{kadavath2022language}, to training a surrogate model to estimate confidence~\cite{shrivastava2023llamas}, to perturbing models to quantify uncertainty~\cite{tanneru2023quantifying}, with some methods leading to more accurate uncertainty estimates than generation probabilities~\cite{johnson2023ru, kuhn2023semantic}. \citet{zhou2023navigating} find that model calibration improves when models are taught to express uncertainty. However, models' \emph{linguistic} uncertainty alone does not necessarily reflect factualness ~\cite{mielke2022reducing, liu2023prudent, tanneru2023quantifying, singh2023confidencecompetence}. For example, a phrase about the weather might have low uncertainty, yet lack factual grounding on current weather data. Relatedly, ~\citet{shrivastava2023llamas} finds that surrogate models trained to estimate confidence can often be better calibrated than the linguistic outputs of generative models. The body of work on uncertainty emission is distinct from, but related to, the idea of self-critiques, in which a model is asked to provide, along with its generated output, a critique of its own or another model or agent's output~\cite{saunders2022selfcritiquing, wu2023autogen}.

\subsection{Highlighting Errors and Regions of Uncertainty}
A few prior works have studied the use of {\em inline highlights} to communicate uncertainty in the generative scenario of AI-powered code completion.  
\citet{ibm} conducted qualitative studies to gain insight into the explainability-related needs of programmers, and found that their design, which highlighted uncertainty in a line of code using a wavy line, revealed the need for: alternative outputs for uncertain code suggestions, explanations for why the AI was uncertain, and the ability to display different levels of uncertainty (e.g., using more colors or hues). Similarly, \citet{vaithilingam-chi2022} conducted qualitative studies and found that programmers found it challenging to read and debug suggestions from Copilot, particularly when its suggestions were long. Their participants expressed a desire for highlighting uncertain code completions. Our work builds on these findings by quantitatively measuring the effect of highlighting uncertainty.

Inline highlights have been used to communicate uncertainty or known errors in many other related domains.
\citet{lank-chi2010} studied the task of handwriting recognition and observed that errors in recognition can pass undetected by the user \cite{lank-chi2010}, but that highlighting the errors can help slow down the user and decrease the error rate.
Static code analysis tools, such as pylint,\footnote{\url{https://www.pylint.org}} have also been popular in software development to improve programmer productivity. These tools are deployed at scale inside IDEs and can, for example, highlight errors in code, such as violations of syntax, coding conventions, or undefined variables. However, they may not catch errors in the logic of the program (e.g., an incorrect regex). Similarly, spell checkers in word processors like MS Word and Google Docs can detect and highlight spelling and common grammatical errors, which can improve incidental learning \cite{lin-rw2017}. AI-based services such as Grammarly can also be used to highlight grammatical and spelling errors in written text.
Our work builds on this finding by incorporating highlighting in our AI-powered code completion tool to improve accuracy and user experience.  \looseness=-1

\subsection{Generative Models and Programmer Performance}

Measuring performance in the programming domain can be challenging, as it is often multi-dimensional and may include subjective elements~\cite{forsgren-acm2021}.
It is difficult to capture programmer productivity in a single metric, and related work has looked at different metrics. For example, 
\citet{al2022readable} found that model-generated code is comparable to code written by people in terms of metrics like complexity and readability, but that people have fewer visual fixations when reading model-generated code. Given that metrics are not all aligned with one another, it is up to designers to determine which metrics to focus on. However, some metrics may lead to unintended outcomes: 
for example, focusing on a metric such as a programmer's usage of AI generated code may be problematic, since \citet{jesse2023large} find that code generation models come with a cost: they are much more likely to output ``simple, stupid bugs,'' so a system that optimizes for this metric might have an increased the presence of these bugs. Another metric may be programmer confidence in code, but, again, we may see some unintended consequences: \citet{perry2022users} find that that programmers are more likely to output insecure code with an AI assistant than without---and with more confidence that they would otherwise. \citet{pudari2023copilot} outline a taxonomy of considerations and current challenges of coding with generative models, including being able to identify and adapt to different coding styles, and suggest design patterns.

Since correctness of code and effort put in by the programmer are key dimensions of performance, in this study, we focus on capturing performance through metrics such as the number of unit tests passed and the time taken. While these metrics may not capture the full complexity of programmer productivity, they do provide valuable insights and can help inform future research on the use of generative models in this context. 

\section{Two Notions of Uncertainty}
\label{sec:preliminary}

In this section, we lay out the two notions of uncertainty that we consider in the study, and the reasoning behind each. We explain how each is calculated and walk through examples of highlights generated using the different notions for the three coding tasks we use in the study. We end with a list of the hypotheses we explore.

\subsection{Uncertainty in Code Generation Models}
\label{sec:uncertainty}

Code generation models are large language models that have been trained or fine-tuned for the task of generating computer code.  A language model is a system that predicts the conditional probability of a token (a character, word, or other string) given either the preceding or surrounding context~\cite{bengio2003neural,radford2019gpt2,bender2021stochastic}. In the case of code generation models, the context may include, for example, prompts, comments, and previously written code. Blocks of recommended code are then generated by sampling from these distributions, one token at a time.  Language models range from simple n-gram models to vastly more sophisticated and expressive models based on modern self-attention architectures like the transformer~\cite{vaswani2017attention}, but all share this common structure. 

The conditional probability of the model producing a particular token in a given context---what we refer to as \emph{generation probability}---can be viewed as a localized notion of the model's uncertainty, and can be directly revealed to end users.  Indeed,  \citet{ibm} propose highlighting low probability lines of code, and OpenAI's online playground interface for their completion API \cite{playground} includes an option to highlight individual tokens with either high or low generation probability. 

While past work has shown generation probabilities to be predictive of which code suggestions programmers are likely to accept~\cite{mozannar2022reading}, this expression of uncertainty may not always line up with programmers' intuition. For example, when naming a new variable, the model will have numerous choices, and this necessarily limits any specific choice's generation probability. However, programmers may incorrectly attribute this uncertainty as indicating a potential error. A similar situation arises when there are multiple correct ways to implement a function.  The disconnect between the meaning of a generation probability and a programmer's intuition may be exacerbated by the fact that language models can be confidently wrong, producing \emph{fluently inadequate}~\cite{martindale2019identifying} outputs that are merely statistically plausible linguistically, but that lose or hallucinate information~\cite[e.g.,][]{hallucinating2022,bender2021stochastic}.

We hypothesize that a more useful notion of token uncertainty to communicate to programmers is the likelihood that they will need modify or delete a generated token in order to arrive at code that meets their needs.  This is not something that can be obtained directly from the code generation model. However, it could potentially be approximated by building a separate \textit{edit model} based on logs of programmers' actions in similar contexts.

In this paper, we learn a closed-world edit model to achieve two goals: (1) to show a proof-of-concept implementation of such an edit model, and (2) as a probe to determine whether revealing such a notion of uncertainty to programmers would enable them to more quickly and accurately produce code when collaborating with a code generation model.  We next describe how we built this model for the purposes of our study.  In Section~\ref{sec:discussion}, we discuss the feasibility of building a more general edit model, suitable for open-world settings.

\subsection{Building the Closed-World Edit Model}
\label{sec:buildingeditmodel}

For the purposes of our study, we did not build a general-purpose edit model, but rather a model that predicts what code programmers would be most likely to edit for the three specific coding tasks and code completions we use in the main study. These tasks are described in detail in the next subsection.  For each of these tasks (as well as two additional tasks that we piloted), we ran the task instructions through OpenAI's Codex model to obtain a completion and corresponding generation probabilities.\footnote{While code completion tools produce output nondeterministically, we used only a single completion for each coding task in our study to reduce variance between participants, ensure the code always contained errors, and enable us to construct the closed-world edit model as described here.}  To generate data for the edit model, we then provided these completions to a set of participants who were given instructions to ``change the completions to ensure they have completed the task properly.'' Participants were also provided with unit tests to check their code. 

We recruited nine participants. As with the main study, all participants were employees of a large technology company located in the United States, and had experience coding in Python.  They were recruited through a mix of direct emails, posts on message boards, and word of mouth, and were paid \$50.  All interviews were conducted over a video-conferencing platform and lasted approximately one hour. Coding tasks were divided such that each was completed by six participants.

We created our closed-world edit model by programmatically tracking which tokens from each completion were edited by participants, considering a token edited if at least one character had been changed, deleted, or commented out. 
For each token in the completion, we then set the probability of an edit to be the fraction of participants who edited it.

\subsection{Coding Tasks and OpenAI's Completions}
\label{sec:codingtasks}

We aimed to select coding tasks that would satisfy the following criteria: (1) they could be completed in about ten minutes by a programmer with some Python experience, and (2) they had reasonable, but imperfect, Codex code completions, with a diversity in error types. 

We initially selected fifteen potential coding tasks from LeetCode~\cite{leetcodeWeb}, a platform to help coders improve their skills and prepare for job interviews, using the ``easy'' setting. 
To generate the code completions for these tasks, we ran the instructions provided by LeetCode as the prompt to Codex (retrieved in June 2022) with the following (mostly default) parameters:
\verb|model:| Code Davinci 002, \verb|temperature:| 0.5, \verb|maximum tokens:| 4000, \verb|top p:|1, \verb|frequency penalty:| 0, \verb|presence penalty:| 0, and \verb|logprobs:| 5. Though the training data for Code Davinci 002 has not been disclosed, and we cannot guarantee it excludes these particular LeetCode problems, selected generations are imperfect, contain errors, and show no evidence suggestive of memorization. After piloting with three participants, we narrowed these down to a set of five tasks, which were used in the edit model training phase, as described above. Finally, we narrowed this down further, selecting the following three tasks that best satisfied our criteria: 

\paragraph{\textsc{Ugly Number}}\footnote{\url{https://leetcode.com/problems/ugly-number/}}
``\emph{An ugly number is a positive integer whose prime factors are limited to 2, 3, and 5. Given an integer \texttt{n}, return \texttt{true} if \texttt{n} is an ugly number.}'' 
The output returned by Codex has many errors, including some syntax that is not in Python (such as ``end'' as a keyword) and the incorrect function name (``\verb|is_ugly|'' instead of ``\verb|isUgly|''). Despite these errors, it has the correct overall logic. For example, the function correctly uses the idea of dividing by 2 until the result is no longer evenly divisible by 2. Because of this, the completion provides useful conceptual help, but a participant would need to correct the many syntax errors to arrive at correct code.

\paragraph{\textsc{Base 7}}\footnote{\url{https://leetcode.com/problems/base-7/}}
``\emph{Given an integer \texttt{num}, return a string of its base 7 representation.}'' 
The output returned by Codex has no syntax errors and only minimal conceptual errors. Specifically, it appends an extra ``0'' to every output. To arrive code that would pass the unit tests, a participant would only need to remove the extra ``0'' from the output.

\paragraph{\textsc{Most Common Word}}\footnote{\url{https://leetcode.com/problems/most-common-word/}}
``\emph{Given a string \texttt{paragraph} and a string array of the banned words \texttt{banned}, return the most frequent word that is not banned. It is guaranteed there is at least one word that is not banned, and that the answer is unique. The words in \texttt{paragraph} are case-insensitive and the answer should be returned in lowercase.}'' 
The output returned by Codex has minimal syntax errors (e.g., using ``List'' which has not been imported, which could be fixed by instead using ``list'' since the non-native Python version is unnecessary). It also has minimal errors in logic, making one simple mistake where a dictionary is edited in place instead of being set equal to a new variable. A participant can arrive at correct code by fixing these two small errors.

\subsection{Setting Thresholds}
\label{sec:thresholds}

In order to highlight the most uncertain tokens, it was necessary to choose thresholds that would determine when tokens are highlighted via generation probabilities and when tokens are highlighted via the edit model. When using generation probabilities, any token with probability \emph{lower} than a specified threshold would be highlighted. Using the edit model, any token with edit probability \emph{higher} than a specified threshold would be highlighted.  To make a fair comparison, we set these thresholds such that the total number of characters highlighted across the three coding tasks was comparable. 

\subsubsection{Edit Model Threshold} Specifically, for the edit model, we chose to highlight tokens that were edited by at least 4 of 6 participants, resulting in a total of 203 highlighted characters across the three tasks. The amount of text highlighted was not too sensitive to this choice; lowering the threshold to 3 participants would yield identical highlights for two tasks, and differ only by a single token in the third.  

\subsubsection{Generation Probabilities Threshold} To set the threshold on generation probabilities, we aimed to highlight a similar number of characters. This could be achieved by highlighting tokens with probability less than 0.694, which would lead to 205 characters highlighted across the three tasks.\footnote{Due to a bug in our code for calculating generation probabilities, our interface failed to highlight 2 tokens, and erroneously highlighted 3 more. Four of those five tokens had generation probability within 0.007 of the 0.694 threshold.
Errors were spread among the three coding tasks and we do not believe they meaningfully impacted the results.}

\begin{figure}[t]
    \centering
    \subfigure[][Ugly Number]{
        \includegraphics[width=0.99\textwidth]{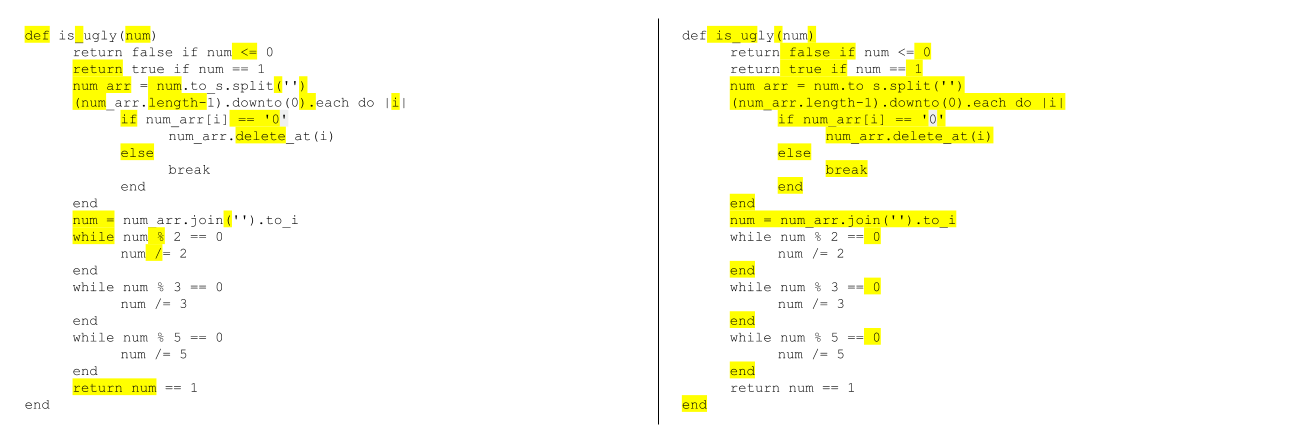}
    }
    \subfigure[][Base 7]{
        \includegraphics[width=0.99\textwidth]{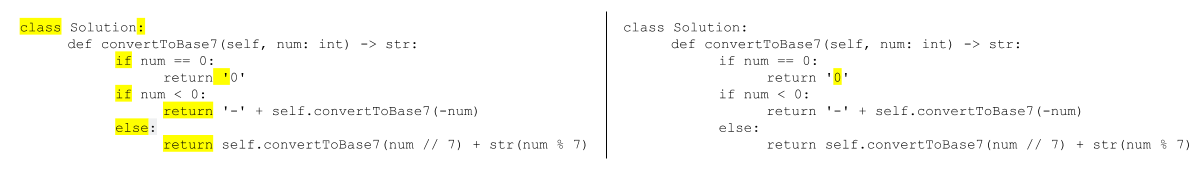} 
    }
    \subfigure[][Most Common Word]{
        \includegraphics[width=0.99\textwidth]{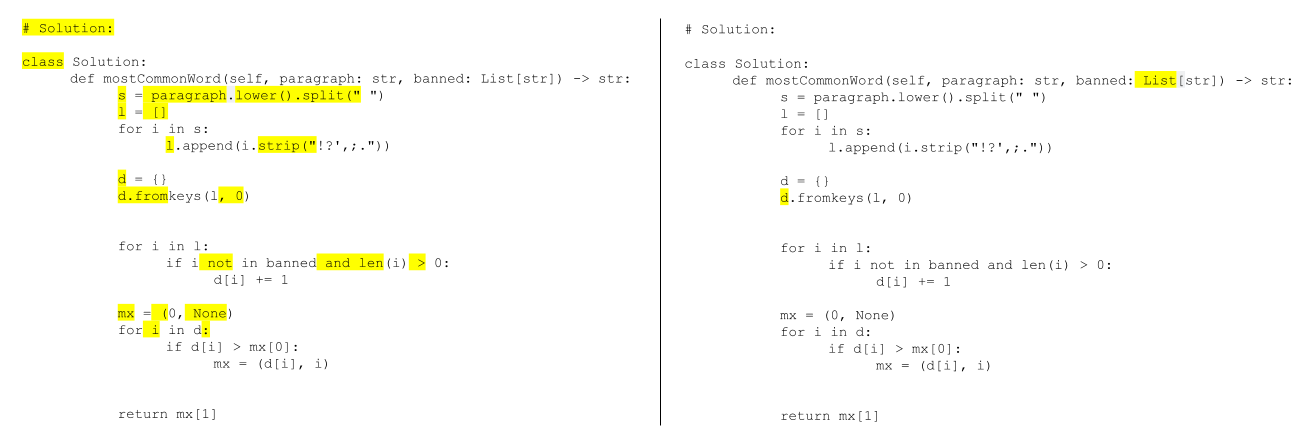}
    }
        \caption{Highlights in the Generation Probability condition (left) and the Edit Model condition (right) for the three coding tasks. The No Highlights condition displays the code completion alone. \label{fig:comparisons}}
\end{figure}

\subsection{Differences and Imperfections in Generation Probabilities and the Edit Model}\label{sec:differences}

Figure~\ref{fig:comparisons} shows the code generations output by Codex for each of the three coding tasks, along with the highlights generated using generation probabilities (left) and the edit model (right). As noted earlier, an approximately equal number of characters are highlighted using the two methods. However, the highlights generated using the edit model concentrate in the completion for the \textsc{Ugly Number} task since participants in the data collection phase for the edit model changed this code the most to get it working.

Overall, with the edit model, highlights tend to correspond to coding errors since they reflect changes performed to initial completions in pursuit of correcting or improving the code.  In contrast, when using generation probabilities, highlights seem to be triggered by that start and end of important sequences (e.g., defining a new class, initiating a new test or loop, or returning a value). Likewise, comparator and Boolean operators (e.g., ``<=,'' ``not,'' ``and'') are often highlighted, perhaps reflecting some uncertainty about the construction and directionality of these expressions. Finally, in several cases, newly introduced variable names are also highlighted. Below we describe in more detail how the highlighted tokens do or do not correspond to errors in the code for each of the three coding tasks:

\begin{itemize}
  
\item For \textsc{Ugly Number}:
As shown in Figure~\ref{fig:comparisons}, highlights obtained using generation probabilities do not capture the more obvious errors, such as having ``end'' as a keyword, not capitalizing ``True'' and ``False,'' and using the wrong function name (``is\_ugly'' instead of ``isUgly,'' as indicated by the prompt). Some tokens are also highlighted that do not contain errors, such as ``def'' at the beginning of the function. In contrast, using the edit model, the aforementioned errors are highlighted, along with a chunk of unnecessary code that should be deleted entirely. 

\item For \textsc{Base 7}:
Using generation probabilities, highlights appear on parts of the code that are correct (e.g., ``class,'' ``if,'' ``else,'' and ``return''). The edit model, on the other hand, leads to a single highlight of the ``0'' which should not appear in the code. Note that simply deleting this ``0'' from the code would be sufficient to pass all unit tests. However, the code would still fail a single edge case not covered in the unit tests. We discuss the implications of this in Section~\ref{sec:timeandacc}. 

\item For \textsc{Most Common Word}:
Here again the highlights obtained using generation probabilities do not focus on the critical parts of the code, but instead highlight a comment that contains ``\# Solution: '' and parts of the code that are correct, such as a list initialization. These highlights do not capture that ``List'' needs to be changed to ``list,'' but do capture the error of ``fromkeys'' being used as a function that modifies variables in place. Using the edit model, on the other hand, only two things are highlighted: the ``List'' and ``fromkeys'' errors. Correcting these errors is, in fact, the minimal change needed to arrive at correct code.

\end{itemize}

\subsection{Pre-registered Hypotheses}
\label{sec:hypotheses}

Since the generation probabilities and the edit model have different distributions and characteristics, we posit that they will similarly have different effects on programmers. Specifically, we pre-registered nine hypotheses,\footnote{Pre-registration link: \href{https://osf.io/tymah}{https://osf.io/tymah} }

which we state informally here; the bolded variables in each hypothesis are defined more formally in Section~\ref{sec:variables}. 

 First, we wanted to explore how much benefit uncertainty highlighting provides to participants in terms of their task performance and efficiency. Ideally, the highlights would  point participants to errors in the code and therefore speed up the process of accurately completing the task. However, highlights may also distract participants, potentially increasing time spent and mistakes made.

\begin{itemize}
\item \textbf{[H1]} Highlight condition will affect the \textbf{time} it takes to complete the coding task. 
\item \textbf{[H2]} Highlight condition will affect \textbf{accuracy} on unit tests.
\end{itemize}

Second, because one goal of working with an AI-powered code completion tool is to reduce the amount of work that programmers need to do, we wanted to explore whether highlighting would affect how much participants have to edit or add to the code. We also examine whether highlighted tokens are more likely to be edited.

\begin{itemize}
\item \textbf{[H3]} Highlight condition will affect the \textbf{number of characters added} to the code.
\item \textbf{[H4]} Highlight condition will affect the \textbf{overall survival rate} of tokens in the code. 
\item \textbf{[H5]} The interaction between highlight condition and whether a given token is highlighted will affect the \textbf{token-level survival rate}.
\end{itemize}

Finally, we wanted to measure participants' subjective preferences for the tools.
We hypothesized that highlights would affect cognitive load, and that participants would feel more favorably towards one condition over the others.

\begin{itemize}
\item \textbf{[H6]} Highlight condition will affect self-reported \textbf{cognitive load}. 
\item \textbf{[H7]} Highlight condition will affect self-reported \textbf{completion utility}. 
\item \textbf{[H8]} Highlight condition will affect self-reported \textbf{highlight utility}. 
\item \textbf{[H9]} Highlight condition will affect self-reported \textbf{rankings} of the code completion tools.
\end{itemize}

\section{Study Design}

\begin{figure}[tb]
    \centering
    \includegraphics[width=0.80\textwidth]{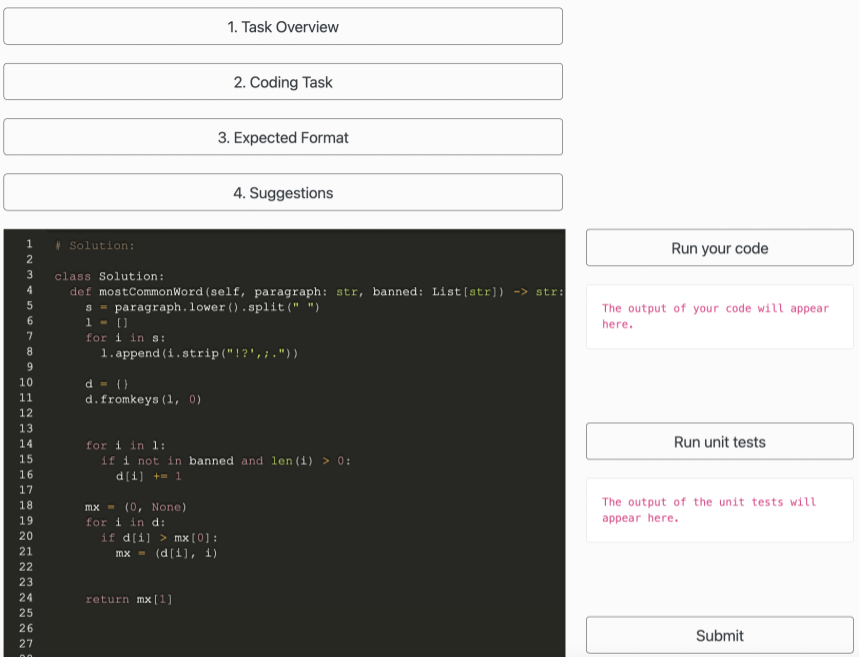}
    \caption{Screenshot of the interface used in our study. The ``Task Overview'' section provides general instructions for the task and interface. The ``Coding Task'' section provides a description of the current coding task. The ``Expected Format'' section describes the code format (e.g., function names and definitions) expected by unit tests. The ``Suggestions'' section provides some suggestions for debugging in our interface. Participants have the option to run their code as a file (``Run your code'') or run our unit tests on their code (``Run unit tests''). When satisfied with their code, they can select ``Submit.''}
    \label{fig:study-interface}
\end{figure}

We conducted a mixed-methods study with 30 participants
consisting of coding tasks and a post-task interview; our sample size was limited to 30 because of practical limitations on interviewer time and participant payments.
All participants were employees of a large technology company located in the United States and had experience coding in Python. None had previously been exposed to our study. Participants were recruited through a mix of direct emails, posts on message boards, and word of mouth. The study was IRB approved with voluntary participation and a payment of \$50. All interviews were conducted over a video-conferencing platform and lasted approximately one hour.  

Of the 30 participants, 15 reported having more than 5 years of experience writing Python code, 11 participants reported having 1--5 years of experience, and 4 reported less than one year of experience. 15 reported that they worked with Python code at least once a week, while 12 reported only using Python ``when needed.'' Comparatively, participants were far less experienced with AI code completion tools: 16 participants never used such a system before. A detailed listing of participant experience is provided in Table~\ref{tab:participants}. 

Finally, among our participants, 5 identified as women, 22 identified as men, and 3 declined to report. 16 were age 24--29, 8 were aged 30--39, 3 were aged 40--49, and 3 declined to report. Additionally, 2 identified as Black, 2 identified as Hispanic/Latino, 5 identified as white, 13 identified as Asian, and 8 declined to report. 

\begin{table}[t!]
\begin{center}
\begin{tabular}{ c|ccc } 
 \hline
 Participant & How long have you been & As of now, how frequently & How often do you use an \\
 & coding in Python? & do you code in Python? & AI-powered code completion
 \\
 & & & tool (e.g., Copilot, AlphaCode)?\\
 \hline
  P1 & 3--5 years & When needed	& Never \\
   P2 & More than 5 years & When needed	& Never \\
   P3 &  1--2 years & Twice a week & Weekly \\
   P4 &  1--2 years & Once a week & Twice a week \\
   P5 &  1--2 years & When needed & When needed \\
   P6 &  < 1 year&Weekly&Never \\
   P7 &  More than 5 years&Daily&Daily \\
   P8 & More than 5 years&Twice a week&Never \\
   P9 &   < 1 year&When needed&Never \\
   P10 &  3--5 years&Daily&Never \\
   P11 &  More than 5 years&Once a week&Never \\
   P12 &  More than 5 years&When needed&When needed \\
   P13 &  More than 5 years&When needed&When needed \\
   P14 &  More than 5 years&Monthly&Daily \\
   P15 &  More than 5 years&Once a week&When needed \\
   P16 &  More than 5 years&Once a week&Never \\
   P17 &  More than 5 years&Daily&When needed \\
   P18 &  More than 5 years&When needed&Never \\
   P19 &  3--5 years&Daily&Weekly \\
   P20 &  3--5 years&Daily&Daily \\
   P21 &  1--2 years&Never&Never \\
   P22 &  3--5 years&When needed&Twice a week \\
   P23&  More than 5 years&Daily&Never \\
   P24 &  3--5 years&When needed&When needed \\
   P25 &  More than 5 years&When needed&Never \\
   P26 &  More than 5 years&Once a week&When needed \\
   P27 &  1--2 years&Once a week&Never \\
   P28 &  < 1 year&When needed&Never \\
   P29 &  < 1 year&When needed&Never \\
   P30 &  More than 5 years&Daily&Never \\
 
\end{tabular}
\caption{Participant details.\label{tab:participants}}
\end{center}
\end{table}

\subsection{Study Procedure}
\label{sec:procedure}

We used a within-subjects design to compare highlighting options. Each participant was asked to complete three coding tasks with three different AI-powered code completion tools. The tools presented the same code completions, and differed only in what was highlighted.
The three highlight conditions we used are as follows:

\begin{itemize}
\item \textbf{No Highlights:} Only the generated code completion was displayed.
\item \textbf{Generation Probability:} The code completion was displayed with highlights on tokens with low generation probability, as in the left side of Figure~\ref{fig:comparisons}. 
\item \textbf{Edit Model:} The code completion was displayed with highlights on tokens with high likelihood 
  of being edited or removed according to the edit model, as in the right side of Figure~\ref{fig:comparisons}. 
\end{itemize}
Both the order of tasks and the assignment of highlight conditions to tasks were randomized.

At the start of the study, participants were given general instructions about the coding tasks they would be asked to complete as well as the interface. 
The interface is shown in Figure~\ref{fig:study-interface}.  Participants were able to view the coding task description along with some examples of inputs and outputs.  They were able to run their code for debugging, and also run a set of provided unit tests on their code at any time. For each coding task, once participants were satisfied with their solution, or after a limit of 10 minutes, they were asked a series of questions rating their experience with the tool and encouraged to explain their responses to the interviewer. 
First, they were asked to rate these statements regarding their subjective perception of the AI tool, each on a 7-point Likert scale 
\begin{enumerate}
\item I found the AI's code completions  helpful as a starting point.
\item (If applicable) I found the AI's highlights helpful in determining what to edit.
\item I would be willing to pay to access the AI's code completions.
\item (If applicable) I would be willing to pay to access the AI's highlights.
\item I found the AI's code completions  distracting. 
\item (If applicable) I found the AI's highlights distracting. 
\end{enumerate}

They were also asked to complete the following reduced\footnote{We reduced the number of TLX questions asked to participants because one of the questions surrounds the idea of physical demand, which we do not think relates meaningfully to our research questions.} NASA RAW TLX (written as ``TLX'' in the rest of the paper) questionnaire~\cite{hart2006nasa}, which aims to measure subjective workload:
\begin{enumerate}
\item How mentally demanding was the task? (Very low to very high; step-size of 20)
\item How hurried or rushed was the pace of the task? (Very low to very high; step-size of 20)
\item How successful were you in accomplishing what you were asked to do? (Prefect to Failure; step-size of 20)
\item How hard did you have to work to accomplish your level of performance? (Very low to very high; step-size of 20)
\item How insecure, discouraged, irritated, stressed, and annoyed were you? (Very low to very high; step-size of 20)
\end{enumerate}

After completing all three coding tasks, participants were prompted by the interface to ``\emph{Please rank the three AIs you used in terms of how satisfied you were with it. Each AI must have a unique rank.}'' The interface also reminded the participants which coding task they completed with each of the three code completion tools.

The study concluded with a semi-structured interview with two parts. In the first part, participants discussed their perception of, and experience with, the three versions of AI tool, including ranking and comparing them. The second part was designed to further explore the design space of uncertainty communication in AI-powered code generation tools. Participants were shown images of five alternative designs to prompt discussion.
We created these designs to explore three main dimensions of the design space: granularity (for instance, uncertainty about a token vs. uncertainty about the whole output),  specificity (for instance, categorical levels of uncertainty vs. precise values), and interactivity (for instance, whether alternative completions should be shown, and whether uncertainty expressions should update automatically as the code is edited). The first two dimensions were informed by the work of \citet{van2019communicating}, who surveyed the psychology and communication literature on uncertainty communication. They distinguished between the effects of varying \textit{what} is being communicated (which relates to granularity) and the \textit{form} of the communication (which relates to specificity). 
The design intended to probe on granularity (Figure~\ref{fig:probe_granularity}) resembles the overall uncertainty scores used in other AI systems, as in~\citet{bansal-chi21},
while the designs intended to probe on specificity (Figure~\ref{fig:probe_specificity}) resemble the types of uncertainty communication implemented in OpenAI's playground~\cite{playground}. The design with alternative completions (Figure~\ref{fig:probe_interactivity}) was also inspired by OpenAI's playground~\cite{playground}.

\subsection{Measured Variables}
\label{sec:variables}

We now formally define the measured variables that we consider in each of the hypotheses stated in Section~\ref{sec:hypotheses}:

\begin{itemize}
    \item \textbf{Time} ({H1}) is measured by the length of time taken to complete the coding task, either by hitting the ``Submit'' button or by reaching the cap of 10 minutes.
    \item \textbf{Accuracy} ({H2}) is measured by the percentage of participant-facing unit tests the participant's code passed for a given coding task. 
    \item \textbf{Number of characters added} ({H3}) is measured by how many new characters the participant added to the code in the end. (This does not include characters that were deleted.) 
    \item \textbf{Overall survival rate} ({H4}) is measured by the aggregate percentage of tokens in the provided code completion that ``survived,'' where survival constitutes not being edited, removed, or commented out. 
    \item \textbf{Token-level survival rate} ({H5}) is measured by whether a given token in the provided code completion survived or not. As before, survival constitutes not being edited, removed, or commented out. 
    \item \textbf{Cognitive load} ({H6}) is measured by the average response to the provided TLX questions, reverse-coded as appropriate.
    \item \textbf{Completion utility} ({H7}) is measured by the average response to questions 1, 3, and 5 (reverse-coded) on subjective perceptions of the tool.
    \item \textbf{Highlight utility} ({H8}) is measured by the average response to questions 2, 4, and 6 (reverse-coded) on subjective perception of the tool.
    \item \textbf{Rank} ({H9}) is measured by the raw rank score provided when the interface prompts the participants to rank the tools at the end of the study, with 1 being best and 3 worst.
\end{itemize}
\section{Quantitative Results}

\begin{table}[t]
\begin{center}
\resizebox{\columnwidth}{!}{
    \begin{tabular}{cccccc } 
     \toprule
     \textbf{Hypothesis} 
&    $\begin{matrix}\textbf{No Highlights}\\ \textbf{(Mean)}\end{matrix}$ 
&    $\begin{matrix}\textbf{Generation}\\ \textbf{Prob (Mean)}\end{matrix}$ 
&    $\begin{matrix}\textbf{Edit Model}\\ \textbf{(Mean)}\end{matrix}$ 
& \textbf{Pairwise Significance} \\
     \midrule
     $\begin{matrix}
        \textbf{H1: Time (Minutes)}
        \\
        p = 0.002\text{**}
     \end{matrix}
     $ 
     & 9.28 & 9.58 & 8.61 & 
     $\begin{matrix}
     \text{Gen Prob > Edit Model**}
     \\
     \text{No Highlights > Edit Model}\dagger
     \end{matrix}
     $
     
     \\     \midrule
     $\begin{matrix}
     \textbf{H2: Accuracy}
     \\
      p = 0.145 
     \end{matrix}$    & 38.3\% & 30.0\% & 50.0\% & --- 
  \\           \midrule
    $\begin{matrix}
    \textbf{H3: Chars Added}
    \\
    p = 0.009\text{**}
    \end{matrix}
    $& 148.9 & 122.4 & 92.2 & 
     No Highlights > Edit Model*
     \\
          \midrule
     $\begin{matrix}
     \textbf{H4: Overall Token}
\\
\textbf{Survival Rate}
     \\
     p = 0.164
     \end{matrix}$
    & 79.3\% & 79.8\% & 75.3\% & --- 
     \\
    \midrule
    $\begin{matrix}
    \textbf{H5: Token-Level}
\\
\textbf{Survival Rate}
    \\
    p < 0.001 \text{***}
    \end{matrix}$
     & 

     \text{Not-HL: 79.3\%}
     & 
     $\begin{matrix}
     \text{HL: 73.9\%}
     \\
     \text{Not-HL: 80.9\%}
     \end{matrix}$
     & 
     $\begin{matrix}
     \text{HL: 35.3\%}
     \\
     \text{Not-HL: 87.1\%}
     \end{matrix}$
     & 
    $ \begin{matrix}
     \text{Edit Model, Not-HL > Edit Model, HL***}
     \\
     \text{Gen Prob, Not-HL > Gen Prob, HL**}
     \\
     \text{Gen Prob, HL > Edit Model, HL***}
     \\
     \text{Edit Model, Not-HL > Gen Prob, Not-HL***}
     \\
     \text{Edit Model, Not-HL > No Highlights***}\end{matrix}$
     \\     
      \midrule
     $\begin{matrix}
     \textbf{H6: Cognitive}
\\
\textbf{Load (0--100)}
     \\ 
     p = 0.228
     \end{matrix}$
     &49.5 & 46.4 & 43.5 & ---
    \\      \midrule
          $\begin{matrix}
          \textbf{H7: Code Completion}
\\
\textbf{Utility (1--7,}
\\
\textbf{higher is better)}
          \\
          p = 0.679
          \end{matrix}$& 4.13 &  4.27 &4.31 & ---
     \\
          \midrule
          $\begin{matrix}
          \textbf{H8: Highlight}
\\
\textbf{Utility (1--7,}
\\
\textbf{higher is better)}
          \\
          p < 0.001 \text{***}
          \end{matrix}$& N/A &  2.93 & 3.92 & Edit Model > Gen Prob**
   \\
    \midrule
     $\begin{matrix}
     \textbf{H9: Rank (1--3, }
\\
\textbf{lower is better)}
     \\
     p = 0.068 \dagger
     \end{matrix}$& 2.03 & 2.10 & 1.87 & Gen Prob > Edit Model$\dagger$
    \end{tabular}
}
\end{center}
\caption{Summary of quantitative results. The left column shows $p$-values obtained via omnibus tests for each hypothesis. The right column shows pairs of conditions that are statistically significantly different or marginally significant. Significance is marked as 
$p < 0.1$ ($\dagger$), $p < 0.05$ (*), $p < 0.01$ (**), or $p < 0.001$ (***).}
\label{tab:overall}
\end{table}

In this section, we analyze each of the pre-registered hypotheses described in Section~\ref{sec:hypotheses}. We provide additional exploratory analyses and participant quotes to shed light on potential reasons why our hypotheses do or do not hold.  Our key quantitative results are summarized in Table~\ref{tab:overall}. 

For each measured variable, unless stated otherwise, we ran a mixed-effects regression model (linear or logistic depending on the data type), using the highlight condition as the fixed-effects independent variable.  Participant ID and coding task were used as random-effects independent variables in all mixed-effect models in this paper, which we omit mentioning for the remaining analyses.  An omnibus $p$-value of the highlight condition was obtained through an ANOVA test, and if significant, we conducted pairwise comparisons with a post-hoc Tukey test.  The left column of Table~\ref{tab:overall} provides $p$-values for the omnibus tests, while the right column shows only the pairs of conditions for which the post-hoc Tukey tests were statistically significant or marginally significant.

All of our analyses are done for all participants (no participants were excluded). In the following section, whenever we subset the data in any way to conduct an analysis (e.g., subsetting for participants who passed all the unit tests), we explicitly say so; otherwise, the analysis was conducted on all participants.

We note that because we randomized both the order in which participants completed coding tasks and the assignment of coding tasks to conditions, the number of participants assigned to each coding task varied by condition, as shown in Table~\ref{tab:numpertask}.  Since we controlled for task assignment in all of our analyses, this randomization should not have a major impact on significance tests.  However, it may have a small impact on the reported means since our sample size is small.

\begin{table}[t]
\begin{center}
    \begin{tabular}{cccc} 
     \toprule
 & No Highlights & Generation Probability & Edit Model \\
\midrule
\textsc{Ugly Number} & 11 & 8 & 11 \\
\textsc{Base 7} & 11 & 10 & 9 \\
 \textsc{Most Common Word} & 8 & 12 & 10 \\
\bottomrule
\end{tabular}
\end{center}
\caption{Number of participants randomly assigned to each highlight condition for each coding task.}
\label{tab:numpertask}
\end{table}

\subsection{Impact of Highlight Condition on Time and Accuracy} \label{sec:timeandacc}

We begin by analyzing the impact of the highlight condition on two quantitative measures of participants' performance: the time it took them to complete the task and their accuracy on unit tests.  We find that participants are significantly faster at completing tasks in the Edit Model condition compared with the Generation Probability condition, even though this effect may be dampened by the 10-minute time limit that we set for each coding task. However, the analysis did not identify statistically significant differences in accuracy. Below we discuss these results in detail.

\paragraph{H1: Highlight condition will affect the \textbf{time} it takes to complete the coding task}

To test hypothesis H1, we used the previously mentioned mixed-effects regression model with time as the dependent variable and highlight condition as the independent variable. 
An omnibus test suggests the effect of highlight condition is significant ($p = 0.002$) and therefore H1 is \emph{supported} by our results.

\begin{figure}[tb]
   \centering
   \includegraphics[width=0.80\textwidth]{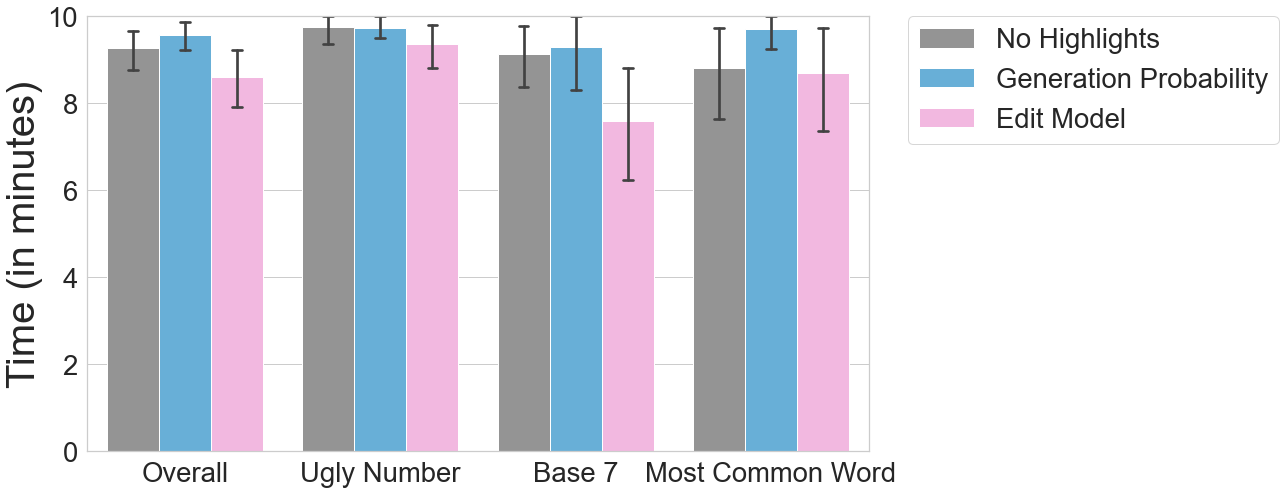}
   \caption{Average time taken per task in minutes (capped at 10) by highlight condition and coding task.}
   \label{fig:H1}
\end{figure}

On average, participants spent $9.27$ minutes per task in the No Highlights condition, $9.61$ minutes in the Generation Probability condition, and $8.59$ minutes in the Edit Model condition.  A post-hoc Tukey test found that the mean value of the time taken was significantly different between the Edit Model condition and the Generation Probability condition ($p = 0.003$), with participants in the Edit Model condition completing their tasks in 89.9\% of the time it took those in the Generation Probability condition on average. There was only a marginally significant difference between the Edit Model condition and the No Highlights condition ($p = 0.066$), with participants completing tasks faster on average in the Edit Model condition. However, as shown in Figure~\ref{fig:H1}, this difference is consistent across tasks.  The analysis identified no significant difference in time taken between the Generation Probability condition and the No Highlights condition ($p = 0.511$).

One potential limitation of our study design is that we capped the amount of time a participant could spend on each coding task at 10 minutes. This was done in order to make time for participants to engage with all three highlight conditions without causing fatigue.  The time constraint affected the distribution of completion times for participants.
In particular, we observed that in both the No Highlights and Generation Probability conditions, $70\%$ of participants were impacted by the 10-minute cutoff, failing to complete the task on time. This number drops to just $44\%$ in the Edit Model condition.  While this makes our results harder to interpret, we conjecture that the difference in the average time taken between the Edit Model condition and the Generation Probability condition would be even more pronounced without the time cap since more participants were affected by the cap in the Generation Probability condition.

\paragraph{H2: Highlight condition will affect \textbf{accuracy} on unit tests}

To test the effect of highlight condition on accuracy, we used a mixed-effects logistic regression model, with the fraction of unit tests passed as the dependent variable. An omnibus test showed the effect of highlight condition was not significant ($p = 0.145$) and therefore H2 is \emph{not supported} by our data.

\begin{figure}[tb]
   \centering
   \includegraphics[width=0.80\textwidth]{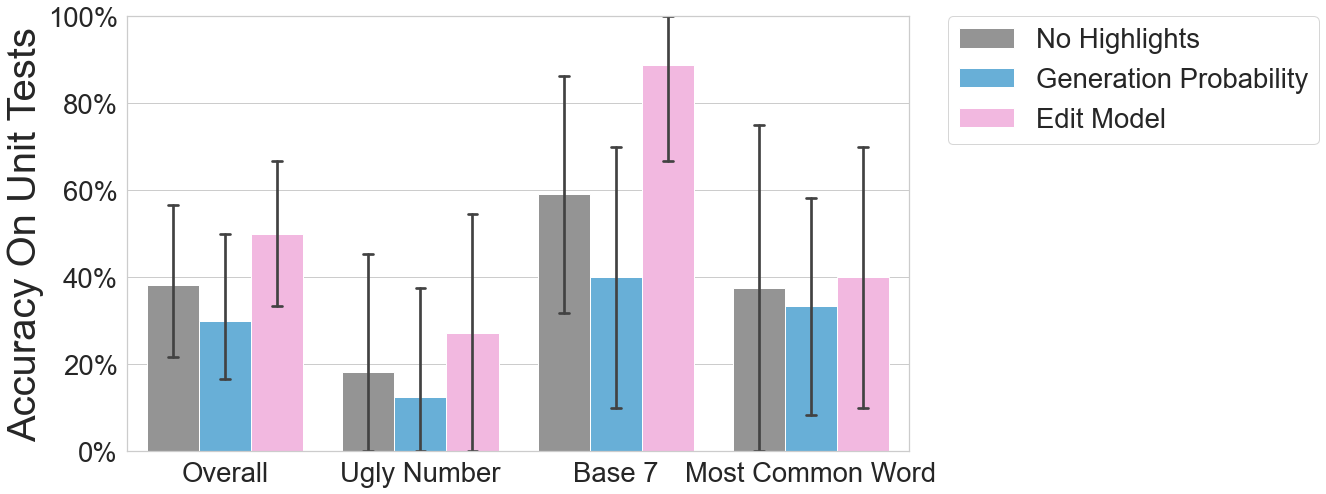}
   \caption{Average accuracy on unit tests by highlight condition and coding task.}
   \label{fig:H2}
\end{figure}

On average, participants achieved $38.3\%$ accuracy in the No Highlights condition, $30.0\%$ accuracy in the Generation Probability condition, and $50.0\%$ accuracy in the Edit Model condition.
Although the analysis did not identify significant differences, we do observe a trend across coding tasks, with participants obtaining  the highest accuracy in the Edit Model condition and the lowest accuracy in the Generation Probability condition across all tasks, as shown in Figure~\ref{fig:H2}.

Our exploratory analyses suggest that this relatively low accuracy for simple coding tasks is at least partly caused by the 10-minute time cap. Specifically, for the \textsc{Most Common Word} and \textsc{Base 7} coding tasks, all participants who completed the task in under 10 minutes had 100\% accuracy on the unit tests across all three conditions. For \textsc{Ugly Number}, participants who completed the task on time in the No Highlights condition had 100\% accuracy, while those in the Generation Probability condition had 33\% accuracy and those in the Edit Model condition had an accuracy of 50\%. 

By editing only the tokens highlighted in the Edit Model condition, participants would be able to pass the provided unit tests for the \textsc{Base 7} coding task, but their code would improperly handle an edge case (because unit tests are not comprehensive). We note that the number of participants who made such a mistake is similar across the Edit Model condition (three out of nine), Generation Probability condition (three out of ten), and the No Highlights condition (two out of eleven), suggesting that this could be an error that programmers are likely to 
make regardless of the presence of highlights. However, participants who made this mistake in the Edit Model condition did submit their code much faster than those in other conditions (6.84 minutes on average, compared with 9.05 minutes in the Generation Probability condition and 7.53 minutes in the No Highlights condition). This observation may suggest a tendency to overrely on the uncertainty highlights for editing code completion.
We discuss this more in Section~\ref{sec:discussion}, but because this analysis is exploratory and inconclusive, we leave this question for future work.

\subsection{Impact of Highlight Condition on Edits Made}

Our second set of hypotheses concerns the specific changes that participants made to the code completions. 
We found that the highlight condition has a significant effect on the number of characters added to the code, with the least characters added in the Edit Model condition.  Interestingly, while the analysis shows no significant effect of highlight condition on overall token survival rate, the specific tokens edited varied by condition: tokens highlighted in the Edit Model condition are not only (already) more likely to be edited by participants, but highlighting them \emph{further boosts} the chance that they will be edited or removed.

\paragraph{H3: Highlight condition will affect the \textbf{number of characters added} to the code}

To test H3, we used the same mixed-effects linear regression model with the number of characters added as the dependent variable.
An omnibus test suggests the effect of highlight condition is significant ($p = 0.009$) and therefore H3 is \emph{supported} by our results.

On average, participants added $148.9$ characters in the No Highlights condition compared with $122.4$ in the Generation Probability condition and only $92.2$ in the Edit Model condition.  However, in a post-hoc Tukey test, only the pairwise difference between the No Highlights condition and the Edit Model condition is significant ($p = 0.010$). If we take characters added as a proxy for (one component of) programmers' effort, this implies that the edit model can reduce programmers' required effort, which is in line with our results on time (H1). 
We acknowledge that this metric can only be considered ``beneficial'' to programmers in conjunction with other metrics, such as accuracy. This is because, in isolation, the metric of edits made could be wrongly optimized to a minimum (e.g., tricking a programmer into thinking an incorrect AI-generated code completion is correct and leading to fewer edits). Because accuracy, for example, is shown to be (albeit not significantly) higher for the Edit Model condition across all tasks, we do not believe that the aforementioned pitfall of optimizing for this metric is the case in this study.

We note that number of characters added only pertains to the number of characters that were left in the code in the end---not those that were added and subsequently removed. Another metric we could have considered to capture ``work'' would be a participant's total number of keystrokes. We examined this in an exploratory analysis and found that the results were similar. The Edit Model condition resulted in the fewest keystrokes, with $209.5$ on average, while there were $237.6$ on average in the Generation Probability condition and $280.1$ in the No Highlights condition.

\paragraph{H4: Highlight condition will affect the \textbf{overall survival rate} of tokens in the code}

To test H4, we used the same mixed-effects logistic regression model with token survival rate as the dependent variable. An omnibus test suggests the effect of highlight condition is not significant ($p = 0.164$) and therefore H4 is \emph{not supported} by our results.

We found that the token survival rate in the No Highlights condition was $79.3\%$, while the rate was $79.8\%$ in the Generation Probability condition, and $75.3\%$ in the Edit Model condition.  Although our analysis does not identify significant differences between these rates, our analysis of H5 below adds nuance to the story, showing that the specific tokens edited \textit{does} vary by highlight condition. 
\begin{figure}[tb]
    \centering
    \subfigure[][]{
        \includegraphics[width=0.48\textwidth]{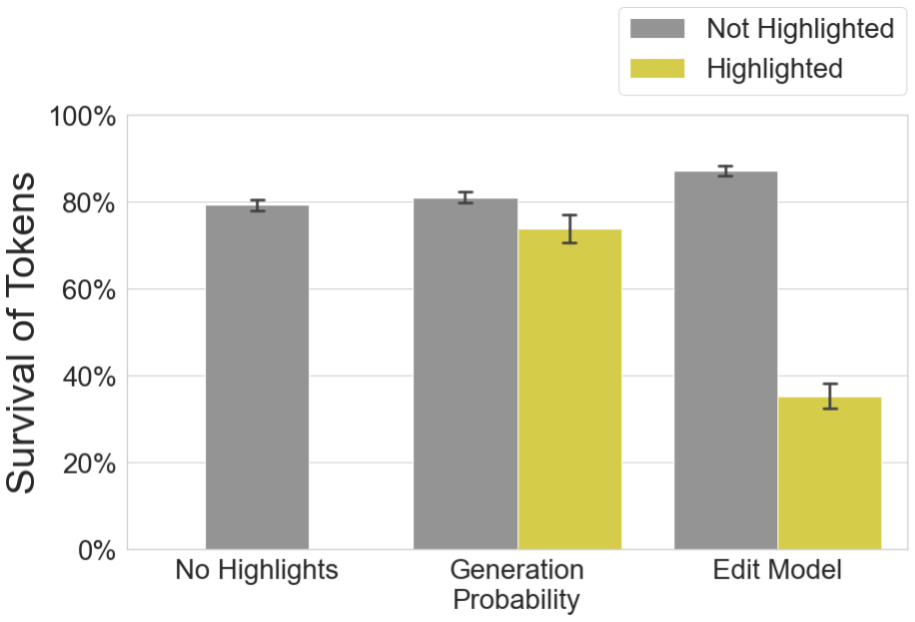}
    }
    \subfigure[][]{
        \includegraphics[width=0.48\textwidth]{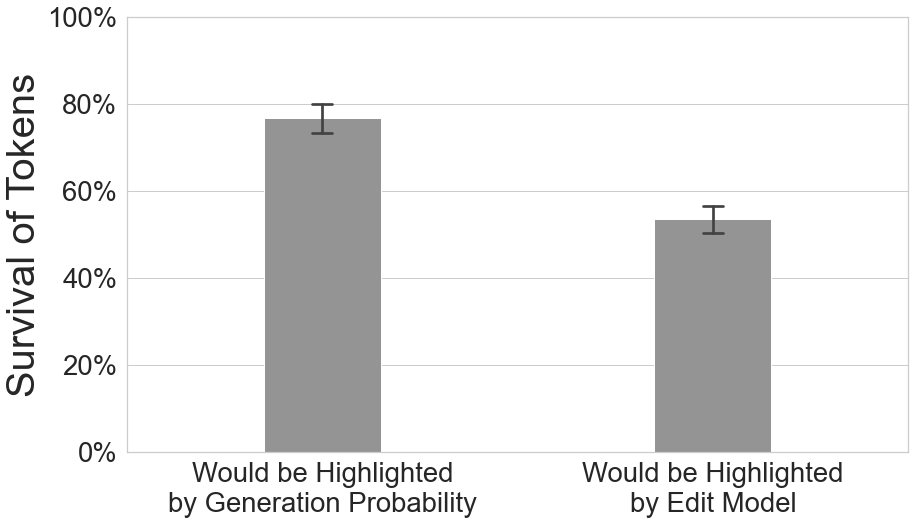}
    }
        \caption{(a): Token-level survival rate for tokens that are highlighted (yellow) or not highlighted (gray) in each of the three experimental conditions.
        (b): Token-level survival rate in the No Highlights condition of tokens that \emph{would have been highlighted} in the Generation Probability condition or Edit Model condition, respectively.}
    \label{fig:H5}
\end{figure}

\paragraph{H5: The interaction between highlight condition and whether a given token is highlighted will affect the \textbf{token-level survival rate}} To test this hypothesis we used a mixed-effects logistic regression model, with token survival rate as the dependent variable, and the interaction between highlight condition and whether a token was highlighted as the fixed-effects independent variables. An omnibus test suggests the effect is significant ($p < 0.001$) and therefore H5 is \emph{supported}.

We observe that for both the Generation Probability condition and the Edit Model condition, tokens that have been highlighted are significantly less likely to survive (i.e., more likely to be edited or removed) than those that have not been highlighted ($p = 0.001$ and $p < 0.001$, respectively). Specifically, for the Generation Probability condition, highlighted tokens have a 73.9\% survival rate, while not-highlighted tokens have a survival rate of 80.9\%.  This difference is even more striking for the Edit Model condition, where highlighted tokens have a survival rate of only 35.3\% compared with 87.1\% for not-highlighted tokens.
Tokens highlighted in the Edit Model condition are significantly less likely to survive than those highlighted in the Generation Probability condition ($p < 0.001$).  
Similarly, the tokens that are \emph{not} highlighted in the Edit Model condition are significantly \emph{more} likely to survive (less likely to be edited or removed) than those not highlighted in the Generation Probability condition ($p < 0.001$).

Additionally, tokens that are not highlighted in the Edit Model condition are more likely to survive than tokens in the No Highlights condition, where, by definition, all tokens are not highlighted (87.1\% chance of survival compared with 79.3\%, $p < 0.001$).
 This effect does not exist for the Generation Probability condition; the analysis identifies no significant difference in whether a token that is not highlighted will survive in the Generation Probability condition compared with a (not highlighted by definition) token in the No Highlights condition. All of these results are summarized in Table~\ref{tab:overall}.

There are two possible factors that could contribute to these results.  One is that highlighting a token causes participants to edit that token more often.  The other is that the highlighted tokens are ones that programmers would be more likely to edit anyway.  Indeed, the tokens highlighted using the edit model are, by definition, those that we would expect programmers to edit most often. To tease apart these factors, we make two additional exploratory comparisons.  First, we limit attention to the set of tokens highlighted in the Edit Model condition and compare the survival rate of \emph{these specific tokens} in the Edit Model condition and the No Highlights condition.  We find that their survival rate is 35.3\% in the Edit Model condition, as reported above, compared with 53.5\% in the No Highlights condition.\footnote{Note that both of these values are higher than we might expect given that tokens were only highlighted if they had a survival rate of no more than 33.3\% based on the original six participants who completed each task. We believe this is due to the small sample size on which the edit model was built and perhaps the presence of the 10-minute time limit.}  This suggests that the effect is coming from a mix of the two factors.  The tokens highlighted in the Edit Model condition are inherently more likely to be edited by programmers, and highlighting them further increases their likelihood of being edited.

Similarly, we can limit attention to the set of tokens highlighted in the Generation Probability condition.  We observe that their survival rate is 73.9\% in the Generation Probability condition, as reported above, compared with 76.8\% in the No Highlights condition. Interestingly, this suggests that the presence of highlights alone is not enough to meaningfully change participants' behavior.  Rather, the highlights must reflect plausible changes that a programmer might be reasonably willing to explore, which generation probabilities do not appear to capture.

\subsection{Impact of Highlight Condition on Cognitive Load and Utility}

Our third set of hypotheses concern self-reported measures of cognitive load and the utility of the completions and highlights. We wanted to measure how much, if at all, our interventions affected participants' preference towards the completion, the highlights, and the code completion tool on the whole. We found that the highlight condition has a significant effect on self-reported utility for the highlights, with higher utility reported in the Edit Model condition compared with the Generation Probability condition. However, the analysis shows no significant effect on cognitive load or code completion utility. Further, we observe that participants' ranking of the tools are dominated by their opinion of the quality of the code completions rather than the highlights.

\paragraph{H6: Highlight condition will affect self-reported \textbf{cognitive load}.}

We defined cognitive load to be the average response across the TLX questions and tested this hypothesis using a mixed-effects linear regression model with cognitive load as the dependent variable. 
An omnibus test shows that the effect of highlight condition is not significant ($p = 0.228$) and therefore H6 is \emph{not supported}.

Participants reported a cognitive load of $49.5$ on average in the No Highlights condition, $46.4$ in the Generation Probability condition, and $43.5$ in the Edit Model condition. Through exploratory analysis, we found that there is a sizable difference in the cognitive load reported by participants who completed a task before the 10-minute mark ($ 35.6$ on average) compared with participants who reached the 10-minute mark ($53.4$). This suggests that whether or not a participant felt they were able to complete the task may have influenced their perceived cognitive load more than the specifics of the highlight condition they were assigned.

\paragraph{H7: Highlight condition will affect self-reported \textbf{code completion utility}.}

Defining code completion utility as in Section~\ref{sec:procedure}, we tested this hypothesis using a mixed-effects linear regression model with code completion utility as the dependent variable.
An omnibus test suggests the effect of highlight condition is not significant ($p = 0.679$) and therefore H7 is \emph{not supported} by our results.

The average code completion utility, on a 1--7 scale with higher being better, was $4.13$ in the No Highlights condition, $4.27$ in the Generation Probability condition, and $4.31$ in the Edit Model condition, with all three of these scores translating roughly as ``neutral'' on the question of code completion utility. The null result reflects that participants judged the utility mainly by the code completion itself (rather than the highlights), which was the same across the three conditions. 


 \begin{figure}[tb]
   \centering
   \includegraphics[width=0.55\textwidth]{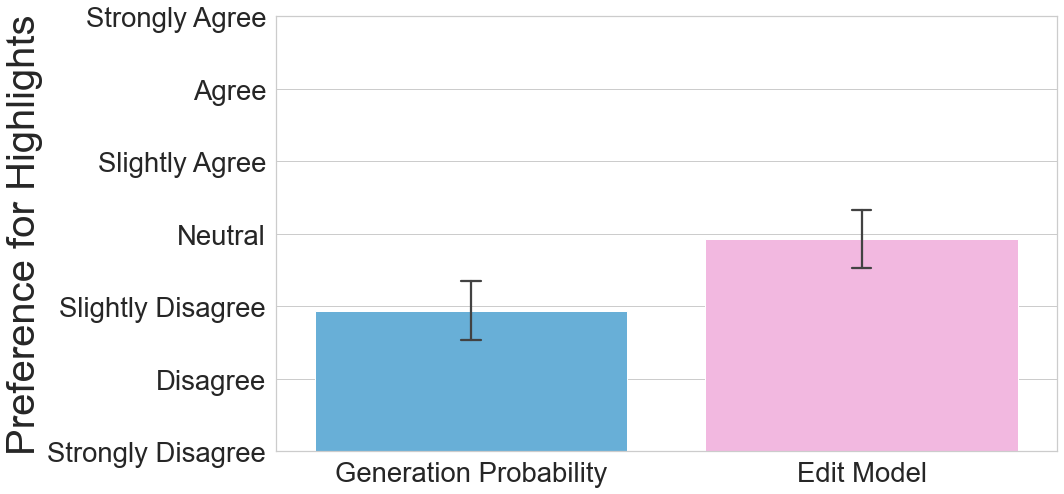}
   \caption{Self-reported highlight utility by condition, where ``agreement'' in this case pertains to the levels of agreement on the utility of highlights on the three questions outlined in Section~\ref{sec:procedure} with the resulting value of Highlight Utility as computed in Section~\ref{sec:variables}.}
   \label{fig:H8}
\end{figure}

\paragraph{H8: Highlight condition will affect self-reported \textbf{highlight utility}.}

Defining highlight utility as in Section~\ref{sec:procedure}, we tested H8 with a mixed-effects linear regression model with highlight utility as the dependent variable.
An omnibus test suggests the effect of highlight condition is significant ($p < 0.001$) and therefore H8 is \emph{supported} by our results.

Specifically, the average highlight utility, on a 1--7 scale with higher being better, was $2.93$ for the Generation Probability condition, which translates roughly towards ``slightly disagree'' on the question of its usefulness, compared with $3.92$ for the Edit Model condition, which translates towards ``neutral'' on the question of its usefulness.
This sizable difference in utility was reflected in participants' comments, which shed more light on the reasons why they preferred the highlights produced with the edit model.  As P4 put it, ``\textit{the highlights in the [Edit Model condition] identified the places where there were compilation or other errors. And so that was particularly useful especially since the editor did not really give good feedback about what the issues were, in particular, when running the unit tests.}''  In contrast, reflecting on the highlights produced using generation probabilities, participants remarked that they were more distracting: “\textit{There were so many there. I think I recall there being one in every two or three lines, if I’m recalling correctly. So it was to the amount that it was noise, so I completely ignored it}'' [P11] and ``\textit{it feels like the highlights in the [Generation Probability condition] were even sort of less useful because they were sort of on stuff that it had gotten right}'' [P18].

Although utility was higher for the Edit Model, average utility was still only ``neutral.'' Our qualitative data sheds some light onto why this might be the case. In particular, participants stated that even though highlights in the Edit Model condition pointed them in the right direction, they didn't know what to do with the information: ``\textit{And the things I don't like is it only highlights the part that is uncertain. So except that there's no other information, so I do not know if I need to delete it or add something to it ...or if the whole line is not correct or if just that word is not correct}'' [P1] and ``\textit{I think it might actually be helpful if I were to, like, hover over the highlight and then it’s telling me something, which probably would help for me to understand what the highlight means}'' [P14].  This is a general limitation of highlighting without including additional information about how code needs to change.

\paragraph{H9: Highlight condition will affect self-reported \textbf{rankings} of the code completion tools.}  
We analyzed this using a cumulative link mixed model\footnote{This analysis (a cumulative link mixed model) differs from the analysis that was pre-registered (a linear mixed-effects model). We chose to make this switch after determining that a cumulative link mixed model is more appropriate for comparing rankings. To start, we note that, normally, we would use a Friedman test for rankings; however, we needed random effects added to our model, due to how influential the question the participant got for a particular condition was in their ranking. Unfortunately, the Friedman test and tests like it (e.g., Wilcox) do not easily permit for random effects. As such, we had to opt for an ordinal regression, wherein the possible outcome values are ordered (as rankings are), as opposed to linear-mixed effects models, which assume the outcome values are normally distributed and continuous. Cumulative link mixed models are ordinal regressions that allow for mixed effects, making it a close ordinal version of the linear-mixed effects model that we had previously pre-registered. } with rank (1 for the best, 2 for the second best, and 3 for the worst) as the dependent variable and highlight condition as the fixed-effects independent variable.  An omnibus test showed the result was only marginally significant ($p = 0.068$) and \textsc{H9} is \emph{marginally supported}.

Participants ranked the code completion tool from the No Highlights condition $2.03$ on average, the tool from the Generation Probability condition $2.10$, and the tool from the Edit Model condition $1.87$ (lower rank means better).  The analysis shows that none of the pairwise differences are significant with our small sample, though the difference between the Generation Probability condition and Edit Model condition is marginally significant ($p = 0.063$).  In total, 9 participants ranked the tool from the No Highlights condition highest, 8 ranked the tool from the Generation Probability condition highest, and the remaining 13 ranked the tool from the Edit Model condition highest. 

Taking this result together with the results for completion utility and highlight utility, we conjecture that participants based their ranking of the tool more on the quality of code completions---which varied between tasks---than the perceived value of the highlights.  (Recall that completions varied by tasks, and tasks were randomly assigned to highlighting conditions.) Indeed, we observed 4 cases in which participants who reported higher highlight utility for the Edit Model condition still ranked the tool from the Generation Probability condition higher, and 2 cases where the reverse was true.  This is in line with participants' comments as they reflected on the code completion tools. For example, P7 stated that ``\textit{the [tool from the Edit Model condition] would be dead last by a long shot... the [tool from the Generation Probability condition] would be number one because it gave me Python code that worked.}''  Similarly, P4 noted that ``\textit{the [tool from the Generation Probability condition] was reasonable code and bad highlights}.''  P18 put it particularly bluntly, saying the tool from the Edit Model condition ``\textit{was just sort of annoying. And, also, I can’t remember if the solution did anything useful when I first read it. And the other two seemed like they were basically already written and running.}''

\section{Further Exploration of the Uncertainty Highlighting Design Space }

Our post-task interview included a discussion of several design probes to further explore the design space of uncertainty highlighting. Our design probes showed alternative designs of uncertainty highlighting that differ in granularity, specificity, and interactivity as described in Section~\ref{sec:procedure}. We analyzed the transcribed content using an inductive approach following our interview structure. We first discuss participants' comments around these three dimensions, then we summarize additional themes regarding participants' preferences regarding uncertainty highlighting and code completion in general. 

\label{apendix:alternatives}
\begin{figure}[t!]
   \centering
   \includegraphics[width=0.50\textwidth]{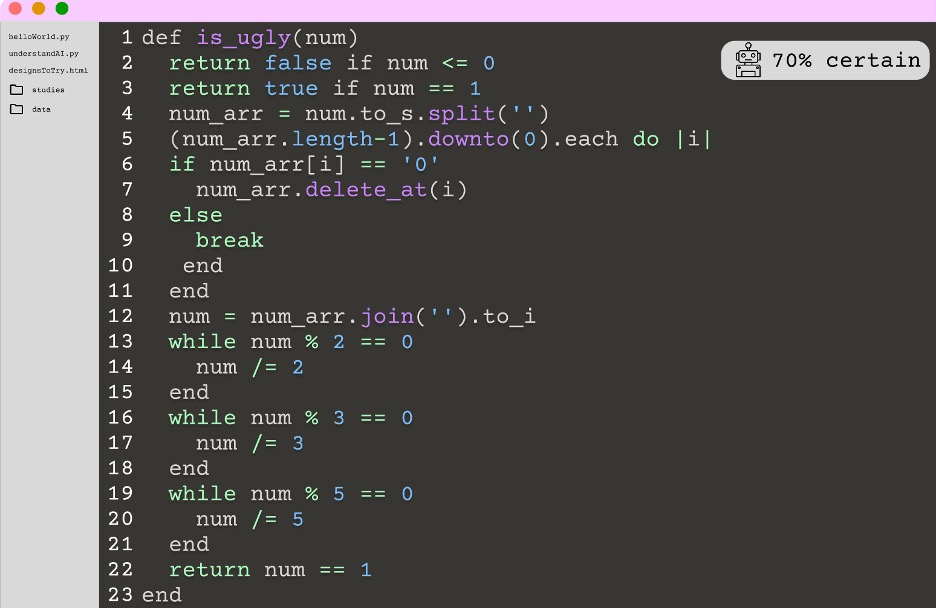}
   \caption{Alternative design that differs in granularity of uncertainty highlighting, where uncertainty is shown as an overall score, rather than per-token.}
   \label{fig:probe_granularity}
\end{figure}

\paragraph{Granularity of uncertainty highlighting} Unlike discriminative models that output a single value for a task, language models output a series of tokens. Because of this, uncertainty about the outputs of language models---and generative models more broadly---can be defined and presented at various levels, from token-level uncertainty (like the generation probabilities and edit model used in our study) to uncertainty about the output as a whole (full code completions, in the case of code completion tools)
and anything in between. 

When being shown an alternative design that displayed uncertainty at the level of the full code completion (Figure~\ref{fig:probe_granularity}), participants generally had negative reactions. Many commented that this method is not granular enough and would not help them find errors. They found that a number without context is abstract, hard to interpret, and doesn't necessarily help them fix or find the errors: ``\textit{running this number like 90 percent or 80 percent, I don't feel too much difference to me}'' [P9] and ``\textit{It's just, if it is not 100 percent correct, we still need to debug it, right?}'' [P12]. Some called out that a single number may bias their overall perception of, and interaction with, the generation: ``\textit{...I think my behavior would depend on the score... if the score is low, I would be pretty frustrated. I would probably just trash the entire thing and rewrite the code myself}'' [P14]. 

On the other hand, this comparison prompted some participants to call out that, while they preferred the token-level uncertainty highlighting, it comes with a tradeoff of potentially highlighting too many things and ``\textit{visually it looks a bit distracting}'' [P14]. Some further commented that with the granular highlighting, the threshold that determines tokens with how much uncertainty to highlight is critical for the user experience: ``\textit{I don't want to see a sea of color in front of me. I'll just assume it thinks it's fine if it's not highlighted}'' [P2].

Some participants offered a middle ground by suggesting the tool could highlight uncertainty at the line level or block level, highlighting digestible blocks of code that are ``\textit{doing a specific task}'' [P8], or ``\textit{if you can chunk out the whole code to separate blocks, and for each separate part give a score then I think that could be helpful}'' [P9].

\begin{figure}[t!]
   \centering
   \includegraphics[width=0.49\textwidth]{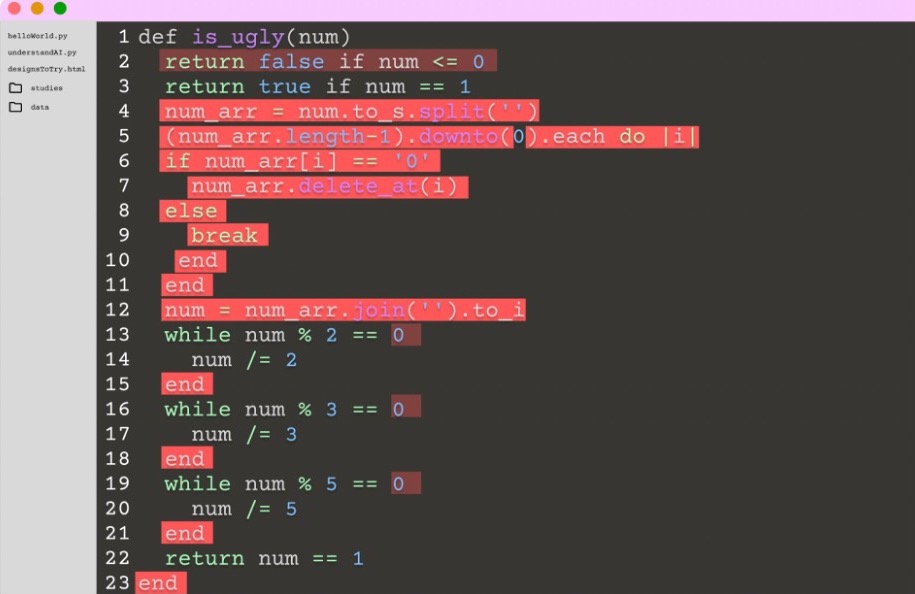}
   \includegraphics[width=0.49\textwidth]{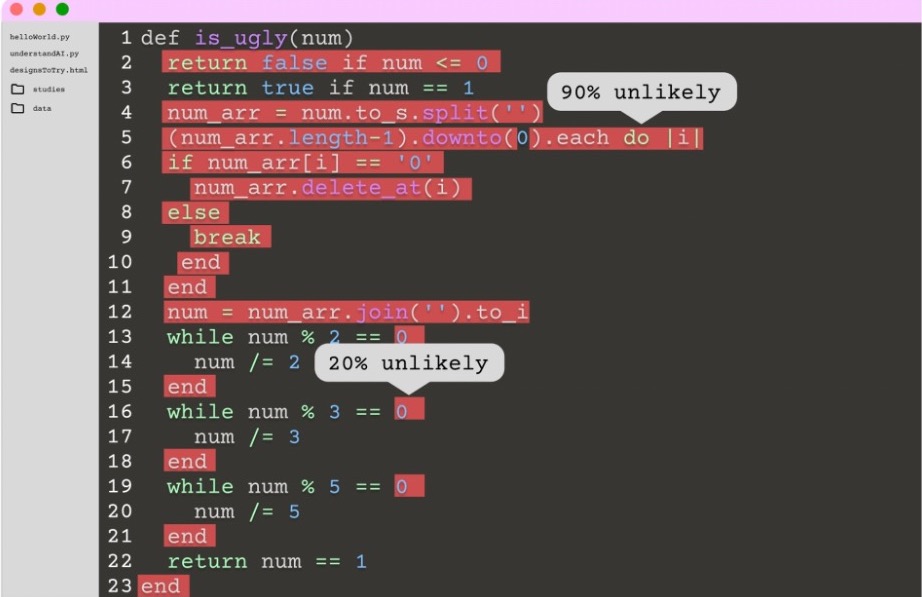}

   \caption{Alternative designs that differ in specificity of uncertainty highlighting, where uncertainty is shown with varying levels of highlight intensity or with the ability to hover over highlights to see exact values.}
   \label{fig:probe_specificity}
\end{figure}

\paragraph{Specificity of Uncertainty Highlighting} We showed two alternative designs to explore the dimension of specificity (Figure~\ref{fig:probe_specificity}). First, in contrast to the single threshold used in our study to determine whether or not a token should be highlighted, one design showed highlights with multiple levels of opacity to convey different levels of uncertainty. The second design revealed the exact numerical quantification of uncertainty (e.g., probability the token would be edited when using the edit model) when hovering over the highlights. Participants almost unanimously pushed back on the idea of displaying numerical values since it is difficult to make meaning out of numbers without context and they do not have the need or capacity to process the additional specificity. For example, P19 commented that ``\textit{I personally think it’s horrible. First of all... I pay very little attention...I wouldn't spend more than, like, say, a minute or even half a minute to interact with the autocompletion... you may know exactly what these numbers are or what exactly these things mean. But, to me, as a user, I wouldn't put in that much of an effort to parse that result... there’s so many things going on in my VS code, I have no idea! My attention’s been drawn everywhere.}''

A number of participants (but not all) preferred the multi-level opacity approach as ``\textit{a nice way of conveying a little bit more information}'' [P2], which can better guide their attention especially working with long code: ``\textit{I’ve got 100 lines of code, I would look at the ones with less opacity first, and then it’ll definitely help me prioritize}'' [P3].

Overall, participants preferred categorical information on uncertainty over being shown the exact quantification of uncertainty. They also highlighted the need to prioritize not overwhelming users as code generation is often used in time-constrained or cognitively demanding settings.  

\begin{figure}[t!]
   \centering
   \includegraphics[width=0.47\textwidth]{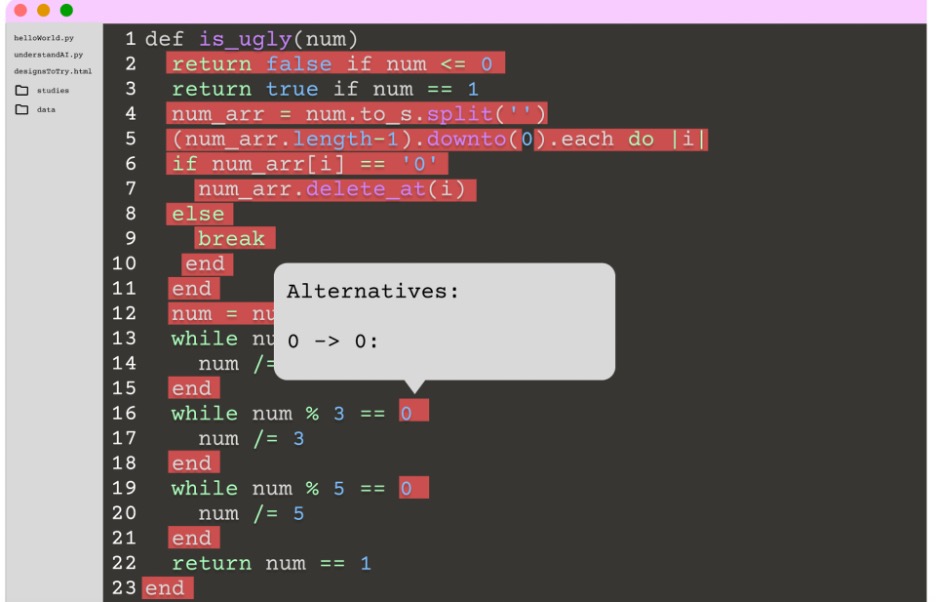}
   \includegraphics[width=0.51\textwidth]{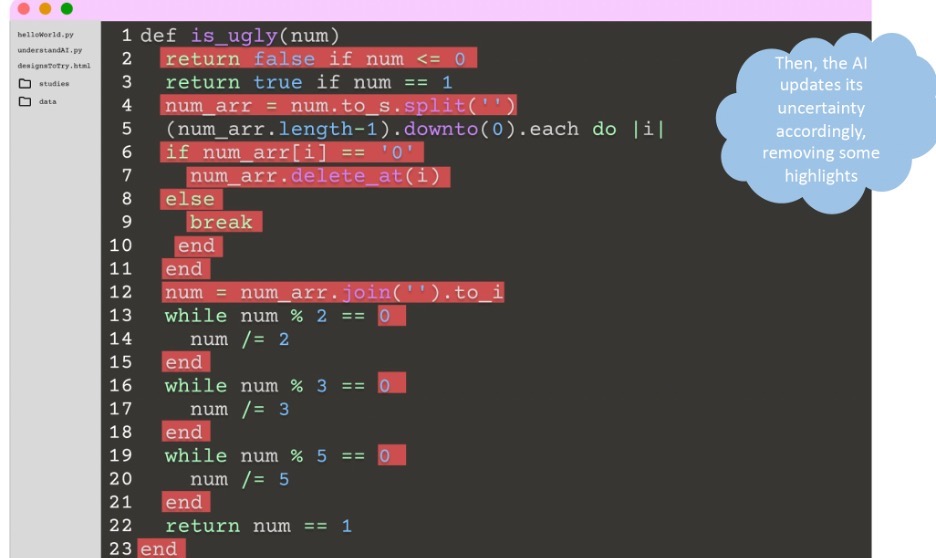}

   \caption{Alternative designs that differ in interactivity of uncertainty highlighting, where uncertainty can yield alternative tokens to generate or changes to the code can update the uncertainty estimations.}
   \label{fig:probe_interactivity}
\end{figure}

\paragraph{Interactivity of Uncertainty Highlighting} We showed participants two example designs with interactivity (Figure~\ref{fig:probe_interactivity}) and encouraged them to come up with other kinds of interactivity they might desire. In one design we showed, alternative tokens were suggested when hovering over the highlights. In the other, the uncertainty highlights were updated as the user made changes to the generated code. Surprisingly, participants pushed back on both ideas. The alternative token suggestion was seen as unhelpful because participants considered token-level changes to be easy fixes much of the time: ``\textit{it's faster for me to go there and insert the colon than it is for me to use, like, let me go to the alternatives}'' [P17]. However, they pointed out that they would find it more useful if appropriate alternative suggestions could happen at line or block level.

Participants generally disliked the idea of the tool taking initiative to auto-update the uncertainty highlights, finding it ``\textit{too confusing and disruptive}'' [P4] or ``\textit{too busy}'' [P15]. They suggested to lower the update frequency so that only significant updates are made, or let the user take the initiative by providing ``\textit{a button to click on}'' [P14] or the ability ``\textit{to toggle ... have them recomputed}'' [P7].

Participants suggested a few other interactivity designs. One feature is to allow users to control the threshold of uncertainty for highlights, such as providing a slider to ``\textit{tune the sensitivity of displaying the highlight}'' [P7] or buttons to choose if ``\textit{you want it to be very sensitive, slightly sensitive, or not sensitive}'' [P16]. Throughout the discussions, many participants requested explanations---not just about how the highlights were computed, but ultimately to diagnose what might have caused this uncertainty. For example, ``\textit{if there's like a clear rule set...clear categories that define and contribute to its level of certainty...to explain why it thinks it's uncertain about this could be useful}'' [P1]. Together, these suggestions reinforce the importance of transparency and user control.

\paragraph{Need to better convey the notion of uncertainty}

A few participants equated uncertainty with errors and found it hard to grasp that the ``\textit{AI was highlighting its own mistakes, which was kind of weird}'' [P15], preferring that ``\textit{if it's not correct, then the code should not be [shown]}'' [P12]. Additional participants found the granularity of the highlight to be potentially misleading, or hard to interpret, as best illustrated by P15's comment: \textit{``when you write something like this, it's tied really closely to what happened before and after and there's a lot of nuances that happen. If the AI is saying like this token might be wrong that basically makes it so the whole function might be wrong because it might be doing something like-- if anything's wrong at this point then like everything after that could be wrong}.'' These comments suggest that people may not have a clear mental model, and could benefit from onboarding information that clearly relates the notion of uncertainty to areas that warrant heightened suspicion (having an elevated \textit{potential} for errors).

\paragraph{Limitations of Highlighting}
Lastly, we find that our participants struggled with the idea that highlighting is not necessarily comprehensive with regards to the potential errors in an AI code completion. Participants cited the fact that highlighting relies on information that is already present in the code completion and will not be able to indicate when an error may be due to missing content: ``...\textit{[this is] an example where highlighting the tokens become kind of tricky because like what was wrong with the solution wasn't necessarily that any part of it was wrong, is that it was missing a case}'' [P14]. This is a limitation of all highlighting techniques, and should be clearly communicated when introducing programmers to such systems, perhaps in tutorials or onboarding materials.

\section{Discussion}
\label{sec:discussion}

Our study finds that highlighting appropriate tokens in generated code can meaningfully impact and improve programmer behavior, and points to a clear path in favor of edit models for this purpose. With the edit model, participants completed tasks faster
while making more targeted edits to the suggested code and reporting higher utility than participants in the Generation Probability condition. Conversely, the Generation Probability condition was often statistically indistinguishable from the baseline No Highlights condition.

Our qualitative findings further suggest that highlighting helps when the highlights are granular, informative, interpretable, and not overwhelming (i.e., when there aren't too many, and participants understand what changes need to be made). Here, a careful balance must be struck: often the preferred design is to do, or show, less rather than more. For example, if model uncertainty is very high, participants preferred suppressing suggestions altogether rather than showing one with many highlights. Likewise, when uncertainty is very low, participants preferred not to see highlights at all, as the base assumption is that only problem areas are flagged (e.g., similar to spell check).

Taken together these findings chart an encouraging, yet cautionary, path for future work in this space. However, there a numerous limitations with this work, and there remain many unknowns. We discuss these limitations and open problems below.

\subsection{Feasibility of Building and Limitations of an Edit Model}
Our work relies on a closed-world edit model, learned on the very set of code completions used in our study. This clearly represents a best-case highly-calibrated scenario. It remains to be demonstrated that we can learn an open-world general-purpose edit model, and that such a model would similarly impact user interaction. Fortunately, systems like GitHub Copilot already consider edits to their code completions as a form of performance metric \cite{Ziegler_2022} and this product-scale data stream could be directly repurposed to this end. As such, the telemetry data collected by these systems, given that they collect initial code generations as well as the edits that people make to those generations, could be adapted to train a model that optimizes for the objective that we've laid out (to predict which tokens people are likely to edit) and have shown with our closed-world model. Notably, in this work, we decided not to train a larger model and rather to use a closed-world model; this was done for study controllability reasons: given that we were unsure whether uncertainty highlighting would have any effects at all, we first wanted to test this in a simpler, more controlled setting with a potentially more accurate model, since this would give signal that is more agnostic to the performance of the edit model. Given that our results are positive and that this data is being collected by code generation companies, we are more optimistic about the utility of constructing an edit model built on larger data.

A crucial limitation of our conception of the edit model, though, is that it relies on the assumption that the data we'd be getting from people---edit data---is representative both of what people want to change for stylistic purposes (e.g., changing a variable name to suit their needs) and of what people want to change for functionality purposes (e.g., changing code that is wrong).
However, it is not clear that people are catching and editing that which they'd otherwise want to change.
For example, if models are good at obscuring mistakes and people never edit those mistakes, then that would not be caught by our edit model. These mistakes that are not edited by people would therefore be left unmarked by a system that relies on an edit model to showcase potential errors. As such, the accuracy of the edit model towards the higher-level objective (to show-case potential errors) is determined by how well the underlying data reflects the assumption we've made that people are able to catch and edit the parts they'd theoretically want to have changed.

\subsection{Representativeness of Tasks, Scenarios, and Participants}
Our study also has limitations with respect to representativeness.  The tasks we include in our study were adapted from LeetCode. These are timed standalone programming challenges, and may not be representative of the day-to-day coding tasks or debugging scenarios that our participants regularly encounter. Likewise, by focusing on unit tests and algorithmic correctness, we may miss other important aspects of code quality such as maintainability and security. Finally, our participants are all professionals in the field of software development, and work for the same US-based technology company. Their experiences may not generalize. For example, students, or those with less experience, may benefit more from the assistance afforded by highlights, or alternatively, may find the lack of explanations more problematic.

Indeed, we observed that experience with Python impacted the results of our study.  We noticed that, qualitatively, participants who were less experienced with Python tended to struggle more in understanding what to do with the highlights and tended to have lower success rates in completing the task across all conditions, and we have confirmed this quantitatively. Aggregating over all experimental conditions, we observe that participants who had less than one year of experience with Python had an average unit test accuracy of 8.3\%, those with one to two years of experience had an average accuracy of 26.6\%, those with three to five years of experience had an average accuracy of 41.6\%, and those with more than five years of experience had an average accuracy of 51.5\%. Future work could explore the differing needs and experiences of expert and novice programmers.

To address these limitations, future work should focus on longitudinal in-situ studies, with broader audiences, and should track outcomes beyond algorithmic correctness. As an example, a future experiment might examine how highlighting strategies impact online metrics such as acceptance rates, or the total proportion of code contributed by the AI system \cite{Ziegler_2022}. Likewise, recent work has suggested that people write less secure code when using such AI systems \cite{perry2022users}, so future work could examine whether highlighting strategies ameliorate this risk. 

\subsection{Impact on Automation Bias}
In the introduction, we motivated this work by presenting the hypothesis that, by highlighting uncertainty, we can help direct human attention to problematic AI generations, and thus reduce automation bias. However, it remains to be demonstrated that the observed benefits (task completion time, targeted edits, preference), translate to increased oversight and decreased automation bias. Nevertheless, we are encouraged that past work has shown that reducing the effort needed to interpret model explanations, or expressions of uncertainty, can increase the likelihood of people overriding AI-induced errors \cite{helena}. We hope to explore these questions in immediate future work.

However, like the code generations themselves, highlights are not perfect, are prone to error, and lack the expressivity needed communicate all classes of error (e.g., errors of omission). In at least one task (\textsc{Base 7}), we observed that an important base case was missing. Participants often attended to other errors flagged by highlights, but submitted their code without addressing the omission. Because a highlighting method cannot highlight code that does not exist, a different design technique may need to be used. For example, annotations in the margins could be used to draw attention to the lines where there are likely to be omissions. Such annotations might be similar to how breakpoints are shown in many IDEs. Likewise, code completion tools such as Copilot already use translucent text in a different font or color to present auto-complete suggestions. It is possible that such text could also be presented in-line where omissions are likely throughout the document, and not just where the text cursor is currently positioned. 

Even if we are able to rectify the issue of highlights not being present due to missing code, there may be other errors of omission, for example when incorrect code is presented but highlights are not. In interviews, several participants mentioned that they interpreted a lack of highlights as signal that the code was correct. Furthermore, even if someone is able to identify the existence of a potential error, they may not be able to correctly determine whether an error truly exists and, if so, correct it. 

Due to these limitations of highlighting and the potential inaccuracies, we may find that we are simply shifting the automation bias such that people are applying an insufficient level of skepticism to the highlights, where before they were insufficiently skeptical of the code completion itself.
\section{Conclusion}
Effective human oversight of AI-powered code completion tools is essential, and requires that programmers be able to efficiently detect and correct generation errors. To this end, prior work---and various commercial systems---use token highlighting to signal which tokens have the lowest likelihood of being generated. However, there have been no empirical studies exploring whether such highlights change programmer behavior in any meaningful way.  

In this paper, we explored whether conveying information about uncertainty via highlights enables programmers to more quickly and accurately produce code when collaborating with an AI-powered code completion tool. Specifically, we considered uncertainty with respect to two distributions: (1) the likelihood of generating the token from the underlying model, and (2) the likelihood of a programmer editing (or deleting) the token once generated. In separate systems, we highlighted tokens that were least likely to be generated, and most likely to be edited, respectively.

Our mixed-methods study with 30 programmers found that highlighting tokens with the highest likelihood of being edited can lead to faster task completion times and more targeted edits, and was subjectively preferred by study participants. In contrast, highlighting tokens according to their generation probabilities did not provide any benefit over a baseline with no highlights.

Further, we explored the design space of token highlighting strategies in post-task interviews with participants. We found significant resistance to designs that highlighted too much, or conveyed too fine-grained a measure of uncertainty. Instead, programmers prefer highlights that are granular, informative, interpretable, and not overwhelming. 
Participants additionally noted that all highlighting strategies are limited because they cannot communicate errors of omission. 

We hope that our positive results, when highlighting tokens most likely to be edited, encourage future and continued work in this important space. At the same time, we hope our results discourage the blind use of generation probabilities for this purpose. Most of all, we hope that research and tooling for human oversight keeps pace with the extraordinarily rapid advancements and releases of code generation models, and generative models more broadly.

\bibliographystyle{ACM-Reference-Format}
\bibliography{references}


\begin{thebibliography}{70}


\ifx \showCODEN    \undefined \def \showCODEN     #1{\unskip}     \fi
\ifx \showDOI      \undefined \def \showDOI       #1{#1}\fi
\ifx \showISBNx    \undefined \def \showISBNx     #1{\unskip}     \fi
\ifx \showISBNxiii \undefined \def \showISBNxiii  #1{\unskip}     \fi
\ifx \showISSN     \undefined \def \showISSN      #1{\unskip}     \fi
\ifx \showLCCN     \undefined \def \showLCCN      #1{\unskip}     \fi
\ifx \shownote     \undefined \def \shownote      #1{#1}          \fi
\ifx \showarticletitle \undefined \def \showarticletitle #1{#1}   \fi
\ifx \showURL      \undefined \def \showURL       {\relax}        \fi
\providecommand\bibfield[2]{#2}
\providecommand\bibinfo[2]{#2}
\providecommand\natexlab[1]{#1}
\providecommand\showeprint[2][]{arXiv:#2}

\bibitem[Al~Madi(2022)]%
        {al2022readable}
\bibfield{author}{\bibinfo{person}{Naser Al~Madi}.} \bibinfo{year}{2022}\natexlab{}.
\newblock \showarticletitle{How readable is model-generated code? examining readability and visual inspection of github copilot}. In \bibinfo{booktitle}{\emph{Proceedings of the 37th IEEE/ACM International Conference on Automated Software Engineering}}. \bibinfo{pages}{1--5}.
\newblock


\bibitem[{Amazon Web Services}(2022)]%
        {codewhisperer_2022}
\bibfield{author}{\bibinfo{person}{{Amazon Web Services}}.} \bibinfo{year}{2022}\natexlab{}.
\newblock \bibinfo{booktitle}{\emph{ML-powered coding companion - Amazon CodeWhisperer}}.
\newblock
\urldef\tempurl%
\url{https://aws.amazon.com/codewhisperer/}
\showURL{%
Retrieved September, 2022 from \tempurl}


\bibitem[Angwin et~al\mbox{.}(2016)]%
        {angwin2016machine}
\bibfield{author}{\bibinfo{person}{Julia Angwin}, \bibinfo{person}{Jeff Larson}, \bibinfo{person}{Surya Mattu}, {and} \bibinfo{person}{Lauren Kirchner}.} \bibinfo{year}{2016}\natexlab{}.
\newblock \showarticletitle{Machine bias: There’s software across the country to predict future criminals and it’s biased against blacks}.
\newblock  (\bibinfo{year}{2016}).
\newblock


\bibitem[Bansal et~al\mbox{.}(2021)]%
        {bansal-chi21}
\bibfield{author}{\bibinfo{person}{Gagan Bansal}, \bibinfo{person}{Tongshuang Wu}, \bibinfo{person}{Joyce Zhou}, \bibinfo{person}{Raymond Fok}, \bibinfo{person}{Besmira Nushi}, \bibinfo{person}{Ece Kamar}, \bibinfo{person}{Marco~Tulio Ribeiro}, {and} \bibinfo{person}{Daniel Weld}.} \bibinfo{year}{2021}\natexlab{}.
\newblock \showarticletitle{Does the whole exceed its parts? the effect of ai explanations on complementary team performance}. In \bibinfo{booktitle}{\emph{Proceedings of the 2021 CHI Conference on Human Factors in Computing Systems}}. \bibinfo{pages}{1--16}.
\newblock


\bibitem[Bender et~al\mbox{.}(2021)]%
        {bender2021stochastic}
\bibfield{author}{\bibinfo{person}{Emily~M. Bender}, \bibinfo{person}{Timnit Gebru}, \bibinfo{person}{Angelina McMillan-Major}, {and} \bibinfo{person}{Shmargaret Shmitchell}.} \bibinfo{year}{2021}\natexlab{}.
\newblock \showarticletitle{On the Dangers of Stochastic Parrots: Can Language Models Be Too Big?}. In \bibinfo{booktitle}{\emph{Proceedings of the 2021 ACM Conference on Fairness, Accountability, and Transparency (FAccT)}}. \bibinfo{pages}{610–623}.
\newblock
\urldef\tempurl%
\url{https://doi.org/10.1145/3442188.3445922}
\showDOI{\tempurl}


\bibitem[Bengio et~al\mbox{.}(2003)]%
        {bengio2003neural}
\bibfield{author}{\bibinfo{person}{Yoshua Bengio}, \bibinfo{person}{R{\'e}jean Ducharme}, \bibinfo{person}{Pascal Vincent}, {and} \bibinfo{person}{Christian Jauvin}.} \bibinfo{year}{2003}\natexlab{}.
\newblock \showarticletitle{A neural probabilistic language model}.
\newblock \bibinfo{journal}{\emph{Journal of Machine Learning Research}} \bibinfo{volume}{3}, \bibinfo{number}{Feb} (\bibinfo{year}{2003}), \bibinfo{pages}{1137--1155}.
\newblock


\bibitem[Bhatt et~al\mbox{.}(2021)]%
        {bhatt2021uncertainty}
\bibfield{author}{\bibinfo{person}{Umang Bhatt}, \bibinfo{person}{Javier Antor{\'a}n}, \bibinfo{person}{Yunfeng Zhang}, \bibinfo{person}{Q~Vera Liao}, \bibinfo{person}{Prasanna Sattigeri}, \bibinfo{person}{Riccardo Fogliato}, \bibinfo{person}{Gabrielle Melan{\c{c}}on}, \bibinfo{person}{Ranganath Krishnan}, \bibinfo{person}{Jason Stanley}, \bibinfo{person}{Omesh Tickoo}, {et~al\mbox{.}}} \bibinfo{year}{2021}\natexlab{}.
\newblock \showarticletitle{Uncertainty as a form of transparency: Measuring, communicating, and using uncertainty}. In \bibinfo{booktitle}{\emph{Proceedings of the 2021 AAAI/ACM Conference on AI, Ethics, and Society}}. \bibinfo{pages}{401--413}.
\newblock


\bibitem[Brown et~al\mbox{.}(2020)]%
        {brown2020language}
\bibfield{author}{\bibinfo{person}{Tom Brown}, \bibinfo{person}{Benjamin Mann}, \bibinfo{person}{Nick Ryder}, \bibinfo{person}{Melanie Subbiah}, \bibinfo{person}{Jared~D Kaplan}, \bibinfo{person}{Prafulla Dhariwal}, \bibinfo{person}{Arvind Neelakantan}, \bibinfo{person}{Pranav Shyam}, \bibinfo{person}{Girish Sastry}, \bibinfo{person}{Amanda Askell}, {et~al\mbox{.}}} \bibinfo{year}{2020}\natexlab{}.
\newblock \showarticletitle{Language models are few-shot learners}.
\newblock \bibinfo{journal}{\emph{Advances in neural information processing systems}}  \bibinfo{volume}{33} (\bibinfo{year}{2020}), \bibinfo{pages}{1877--1901}.
\newblock


\bibitem[Cai et~al\mbox{.}(2019)]%
        {cai2019hello}
\bibfield{author}{\bibinfo{person}{Carrie~J Cai}, \bibinfo{person}{Samantha Winter}, \bibinfo{person}{David Steiner}, \bibinfo{person}{Lauren Wilcox}, {and} \bibinfo{person}{Michael Terry}.} \bibinfo{year}{2019}\natexlab{}.
\newblock \showarticletitle{" Hello AI": Uncovering the Onboarding Needs of Medical Practitioners for Human-AI Collaborative Decision-Making}.
\newblock \bibinfo{journal}{\emph{Proceedings of the ACM on Human-computer Interaction}} \bibinfo{volume}{3}, \bibinfo{number}{CSCW} (\bibinfo{year}{2019}), \bibinfo{pages}{1--24}.
\newblock


\bibitem[Cai et~al\mbox{.}(2021)]%
        {cai-chi2021}
\bibfield{author}{\bibinfo{person}{Carrie~J. Cai}, \bibinfo{person}{Samantha Winter}, \bibinfo{person}{David~F. Steiner}, \bibinfo{person}{Lauren Wilcox}, {and} \bibinfo{person}{Michael Terry}.} \bibinfo{year}{2021}\natexlab{}.
\newblock \showarticletitle{Onboarding Materials as Cross-functional Boundary Objects for Developing AI Assistants}.
\newblock \bibinfo{journal}{\emph{Extended Abstracts of the 2021 CHI Conference on Human Factors in Computing Systems}} (\bibinfo{year}{2021}).
\newblock


\bibitem[Chakrabarty et~al\mbox{.}(2022)]%
        {chakrabarty-arxiv2022}
\bibfield{author}{\bibinfo{person}{Tuhin Chakrabarty}, \bibinfo{person}{Vishakh Padmakumar}, {and} \bibinfo{person}{He He}.} \bibinfo{year}{2022}\natexlab{}.
\newblock \showarticletitle{Help me write a poem: Instruction Tuning as a Vehicle for Collaborative Poetry Writing}.
\newblock \bibinfo{journal}{\emph{arXiv preprint arXiv:2210.13669}} (\bibinfo{year}{2022}).
\newblock


\bibitem[Clark et~al\mbox{.}(2018)]%
        {clark-iui2018}
\bibfield{author}{\bibinfo{person}{Elizabeth Clark}, \bibinfo{person}{Anne~Spencer Ross}, \bibinfo{person}{Chenhao Tan}, \bibinfo{person}{Yangfeng Ji}, {and} \bibinfo{person}{Noah~A Smith}.} \bibinfo{year}{2018}\natexlab{}.
\newblock \showarticletitle{Creative writing with a machine in the loop: Case studies on slogans and stories}. In \bibinfo{booktitle}{\emph{23rd International Conference on Intelligent User Interfaces}}. \bibinfo{pages}{329--340}.
\newblock


\bibitem[{DeepMind}(2022)]%
        {alphacode_2022}
\bibfield{author}{\bibinfo{person}{{DeepMind}}.} \bibinfo{year}{2022}\natexlab{}.
\newblock \bibinfo{booktitle}{\emph{AlphaCode}}.
\newblock
\urldef\tempurl%
\url{https://alphacode.deepmind.com/}
\showURL{%
Retrieved September, 2022 from \tempurl}


\bibitem[Dhami and Mandel(2022)]%
        {dhami-elsevier2022}
\bibfield{author}{\bibinfo{person}{Mandeep~K Dhami} {and} \bibinfo{person}{David~R Mandel}.} \bibinfo{year}{2022}\natexlab{}.
\newblock \showarticletitle{Communicating uncertainty using words and numbers}.
\newblock \bibinfo{journal}{\emph{Trends in Cognitive Sciences}} (\bibinfo{year}{2022}).
\newblock


\bibitem[Forsgren et~al\mbox{.}(2021)]%
        {forsgren-acm2021}
\bibfield{author}{\bibinfo{person}{Nicole Forsgren}, \bibinfo{person}{Margaret-Anne Storey}, \bibinfo{person}{Chandra Maddila}, \bibinfo{person}{Thomas Zimmermann}, \bibinfo{person}{Brian Houck}, {and} \bibinfo{person}{Jenna Butler}.} \bibinfo{year}{2021}\natexlab{}.
\newblock \showarticletitle{The SPACE of Developer Productivity: There's more to it than you think.}
\newblock \bibinfo{journal}{\emph{Queue}} \bibinfo{volume}{19}, \bibinfo{number}{1} (\bibinfo{year}{2021}), \bibinfo{pages}{20--48}.
\newblock


\bibitem[{GitHub}(2022)]%
        {copilot_2022}
\bibfield{author}{\bibinfo{person}{{GitHub}}.} \bibinfo{year}{2022}\natexlab{}.
\newblock \bibinfo{booktitle}{\emph{GitHub Copilot - Your AI pair programmer}}.
\newblock
\urldef\tempurl%
\url{https://github.com/features/copilot/}
\showURL{%
Retrieved September, 2022 from \tempurl}


\bibitem[Gonzalez et~al\mbox{.}(2021)]%
        {gonzalez-acl2021}
\bibfield{author}{\bibinfo{person}{Ana~Valeria Gonzalez}, \bibinfo{person}{Gagan Bansal}, \bibinfo{person}{Angela Fan}, \bibinfo{person}{Yashar Mehdad}, \bibinfo{person}{Robin Jia}, {and} \bibinfo{person}{Srini Iyer}.} \bibinfo{year}{2021}\natexlab{}.
\newblock \showarticletitle{Do Explanations Help Users Detect Errors in Open-Domain QA? An Evaluation of Spoken vs. Visual Explanations}. In \bibinfo{booktitle}{\emph{Findings of ACL}}.
\newblock


\bibitem[Green and Chen(2020)]%
        {green-chi2020}
\bibfield{author}{\bibinfo{person}{Ben Green} {and} \bibinfo{person}{Yiling Chen}.} \bibinfo{year}{2020}\natexlab{}.
\newblock \showarticletitle{Algorithmic Risk Assessments Can Alter Human Decision-Making Processes in High-Stakes Government Contexts}.
\newblock \bibinfo{journal}{\emph{Proceedings of the ACM on Human-Computer Interaction}}  \bibinfo{volume}{5} (\bibinfo{year}{2020}), \bibinfo{pages}{1 -- 33}.
\newblock


\bibitem[Guo et~al\mbox{.}(2017)]%
        {guo-pmlr2017}
\bibfield{author}{\bibinfo{person}{Chuan Guo}, \bibinfo{person}{Geoff Pleiss}, \bibinfo{person}{Yu Sun}, {and} \bibinfo{person}{Kilian~Q Weinberger}.} \bibinfo{year}{2017}\natexlab{}.
\newblock \showarticletitle{On calibration of modern neural networks}. In \bibinfo{booktitle}{\emph{International conference on machine learning}}. PMLR, \bibinfo{pages}{1321--1330}.
\newblock


\bibitem[Hart(2006)]%
        {hart2006nasa}
\bibfield{author}{\bibinfo{person}{Sandra~G Hart}.} \bibinfo{year}{2006}\natexlab{}.
\newblock \showarticletitle{NASA-task load index (NASA-TLX); 20 years later}. In \bibinfo{booktitle}{\emph{Proceedings of the human factors and ergonomics society annual meeting}}, Vol.~\bibinfo{volume}{50}. Sage publications Sage CA: Los Angeles, CA, \bibinfo{pages}{904--908}.
\newblock


\bibitem[Hayashi and Wakabayashi(2017)]%
        {hayashi2017can}
\bibfield{author}{\bibinfo{person}{Yugo Hayashi} {and} \bibinfo{person}{Kosuke Wakabayashi}.} \bibinfo{year}{2017}\natexlab{}.
\newblock \showarticletitle{Can AI become reliable source to support human decision making in a court scene?}. In \bibinfo{booktitle}{\emph{Companion of the 2017 ACM Conference on Computer Supported Cooperative Work and Social Computing}}. \bibinfo{pages}{195--198}.
\newblock


\bibitem[Jacobs et~al\mbox{.}(2021)]%
        {jacobs-psych2021}
\bibfield{author}{\bibinfo{person}{Maia~L. Jacobs}, \bibinfo{person}{Melanie~Fernandes Pradier}, \bibinfo{person}{Thomas~H. McCoy}, \bibinfo{person}{Roy~H. Perlis}, \bibinfo{person}{Finale Doshi-Velez}, {and} \bibinfo{person}{Krzysztof~Z Gajos}.} \bibinfo{year}{2021}\natexlab{}.
\newblock \showarticletitle{How machine-learning recommendations influence clinician treatment selections: the example of antidepressant selection}.
\newblock \bibinfo{journal}{\emph{Translational Psychiatry}}  \bibinfo{volume}{11} (\bibinfo{year}{2021}).
\newblock


\bibitem[Jesse et~al\mbox{.}(2023)]%
        {jesse2023large}
\bibfield{author}{\bibinfo{person}{Kevin Jesse}, \bibinfo{person}{Toufique Ahmed}, \bibinfo{person}{Premkumar~T. Devanbu}, {and} \bibinfo{person}{Emily Morgan}.} \bibinfo{year}{2023}\natexlab{}.
\newblock \bibinfo{title}{Large Language Models and Simple, Stupid Bugs}.
\newblock
\newblock
\showeprint[arxiv]{2303.11455}~[cs.SE]


\bibitem[Ji et~al\mbox{.}(2022)]%
        {hallucinating2022}
\bibfield{author}{\bibinfo{person}{Ziwei Ji}, \bibinfo{person}{Nayeon Lee}, \bibinfo{person}{Rita Frieske}, \bibinfo{person}{Tiezheng Yu}, \bibinfo{person}{Dan Su}, \bibinfo{person}{Yan Xu}, \bibinfo{person}{Etsuko Ishii}, \bibinfo{person}{Yejin Bang}, \bibinfo{person}{Andrea Madotto}, {and} \bibinfo{person}{Pascale Fung}.} \bibinfo{year}{2022}\natexlab{}.
\newblock \showarticletitle{Survey of Hallucination in Natural Language Generation}.
\newblock \bibinfo{journal}{\emph{Comput. Surveys}} (\bibinfo{date}{nov} \bibinfo{year}{2022}).
\newblock
\urldef\tempurl%
\url{https://doi.org/10.1145/3571730}
\showDOI{\tempurl}


\bibitem[Jiang et~al\mbox{.}(2021)]%
        {jiang-tacl2021}
\bibfield{author}{\bibinfo{person}{Zhengbao Jiang}, \bibinfo{person}{Jun Araki}, \bibinfo{person}{Haibo Ding}, {and} \bibinfo{person}{Graham Neubig}.} \bibinfo{year}{2021}\natexlab{}.
\newblock \showarticletitle{How can we know when language models know? on the calibration of language models for question answering}.
\newblock \bibinfo{journal}{\emph{Transactions of the Association for Computational Linguistics}}  \bibinfo{volume}{9} (\bibinfo{year}{2021}), \bibinfo{pages}{962--977}.
\newblock


\bibitem[Johnson et~al\mbox{.}(2023)]%
        {johnson2023ru}
\bibfield{author}{\bibinfo{person}{Daniel~D Johnson}, \bibinfo{person}{Daniel Tarlow}, {and} \bibinfo{person}{Christian Walder}.} \bibinfo{year}{2023}\natexlab{}.
\newblock \showarticletitle{RU-SURE? Uncertainty-Aware Code Suggestions By Maximizing Utility Across Random User Intents}.
\newblock \bibinfo{journal}{\emph{arXiv preprint arXiv:2303.00732}} (\bibinfo{year}{2023}).
\newblock


\bibitem[Kadavath et~al\mbox{.}(2022)]%
        {kadavath2022language}
\bibfield{author}{\bibinfo{person}{Saurav Kadavath}, \bibinfo{person}{Tom Conerly}, \bibinfo{person}{Amanda Askell}, \bibinfo{person}{Tom Henighan}, \bibinfo{person}{Dawn Drain}, \bibinfo{person}{Ethan Perez}, \bibinfo{person}{Nicholas Schiefer}, \bibinfo{person}{Zac Hatfield-Dodds}, \bibinfo{person}{Nova DasSarma}, \bibinfo{person}{Eli Tran-Johnson}, {et~al\mbox{.}}} \bibinfo{year}{2022}\natexlab{}.
\newblock \showarticletitle{Language models (mostly) know what they know}.
\newblock \bibinfo{journal}{\emph{arXiv preprint arXiv:2207.05221}} (\bibinfo{year}{2022}).
\newblock


\bibitem[Kalliamvakou(2022)]%
        {copilot-gh2022}
\bibfield{author}{\bibinfo{person}{Eirini Kalliamvakou}.} \bibinfo{year}{2022}\natexlab{}.
\newblock \bibinfo{booktitle}{\emph{Research: quantifying GitHub Copilot’s impact on developer productivity and happiness}}.
\newblock
\urldef\tempurl%
\url{https://github.blog/2022-09-07-research-quantifying-github-copilots-impact-on-developer-productivity-and-happiness/}
\showURL{%
\tempurl}


\bibitem[Khakhar et~al\mbox{.}(2023)]%
        {khakhar2023pac}
\bibfield{author}{\bibinfo{person}{Adam Khakhar}, \bibinfo{person}{Stephen Mell}, {and} \bibinfo{person}{Osbert Bastani}.} \bibinfo{year}{2023}\natexlab{}.
\newblock \bibinfo{title}{PAC Prediction Sets for Large Language Models of Code}.
\newblock
\newblock
\showeprint[arxiv]{2302.08703}~[cs.LG]


\bibitem[Kocielnik et~al\mbox{.}(2019)]%
        {kocielnik-chi2019}
\bibfield{author}{\bibinfo{person}{Rafal Kocielnik}, \bibinfo{person}{Saleema Amershi}, {and} \bibinfo{person}{Paul~N. Bennett}.} \bibinfo{year}{2019}\natexlab{}.
\newblock \showarticletitle{Will You Accept an Imperfect AI?: Exploring Designs for Adjusting End-user Expectations of AI Systems}.
\newblock \bibinfo{journal}{\emph{Proceedings of the 2019 CHI Conference on Human Factors in Computing Systems}} (\bibinfo{year}{2019}).
\newblock


\bibitem[Kotelanski et~al\mbox{.}(2023)]%
        {kotelanski2023methods}
\bibfield{author}{\bibinfo{person}{Maia Kotelanski}, \bibinfo{person}{Robert Gallo}, \bibinfo{person}{Ashwin Nayak}, {and} \bibinfo{person}{Thomas Savage}.} \bibinfo{year}{2023}\natexlab{}.
\newblock \bibinfo{title}{Methods to Estimate Large Language Model Confidence}.
\newblock
\newblock
\showeprint[arxiv]{2312.03733}~[cs.CL]


\bibitem[Kuhn et~al\mbox{.}(2023)]%
        {kuhn2023semantic}
\bibfield{author}{\bibinfo{person}{Lorenz Kuhn}, \bibinfo{person}{Yarin Gal}, {and} \bibinfo{person}{Sebastian Farquhar}.} \bibinfo{year}{2023}\natexlab{}.
\newblock \showarticletitle{Semantic uncertainty: Linguistic invariances for uncertainty estimation in natural language generation}.
\newblock \bibinfo{journal}{\emph{arXiv preprint arXiv:2302.09664}} (\bibinfo{year}{2023}).
\newblock


\bibitem[Lai et~al\mbox{.}(2020)]%
        {lai-chi2020}
\bibfield{author}{\bibinfo{person}{Vivian Lai}, \bibinfo{person}{Han Liu}, {and} \bibinfo{person}{Chenhao Tan}.} \bibinfo{year}{2020}\natexlab{}.
\newblock \showarticletitle{"Why is 'Chicago' deceptive?" Towards Building Model-Driven Tutorials for Humans}.
\newblock \bibinfo{journal}{\emph{Proceedings of the 2020 CHI Conference on Human Factors in Computing Systems}} (\bibinfo{year}{2020}).
\newblock


\bibitem[Lank et~al\mbox{.}(2010)]%
        {lank-chi2010}
\bibfield{author}{\bibinfo{person}{Edward Lank}, \bibinfo{person}{Ryan Stedman}, {and} \bibinfo{person}{Michael Terry}.} \bibinfo{year}{2010}\natexlab{}.
\newblock \showarticletitle{Estimating residual error rate in recognized handwritten documents using artificial error injection}. In \bibinfo{booktitle}{\emph{Proceedings of the SIGCHI Conference on Human Factors in Computing Systems}}. \bibinfo{pages}{1--4}.
\newblock


\bibitem[LeetCode(2015)]%
        {leetcodeWeb}
\bibfield{author}{\bibinfo{person}{LeetCode}.} \bibinfo{year}{2015}\natexlab{}.
\newblock \showarticletitle{The world's leading online programming learning platform}.
\newblock
\urldef\tempurl%
\url{https://leetcode.com/}
\showURL{%
\tempurl}


\bibitem[Liem et~al\mbox{.}(2018)]%
        {liem_2018}
\bibfield{author}{\bibinfo{person}{Cynthia~CS Liem}, \bibinfo{person}{Markus Langer}, \bibinfo{person}{Andrew Demetriou}, \bibinfo{person}{Annemarie~MF Hiemstra}, \bibinfo{person}{Achmadnoer~Sukma Wicaksana}, \bibinfo{person}{Marise~Ph Born}, {and} \bibinfo{person}{Cornelius~J K{\"o}nig}.} \bibinfo{year}{2018}\natexlab{}.
\newblock \showarticletitle{Psychology Meets Machine Learning: Interdisciplinary Perspectives on Algorithmic Job Candidate Screening}.
\newblock In \bibinfo{booktitle}{\emph{Explainable and Interpretable Models in Computer Vision and Machine Learning}}. \bibinfo{publisher}{Springer}, \bibinfo{pages}{197--253}.
\newblock


\bibitem[Lin et~al\mbox{.}(2017)]%
        {lin-rw2017}
\bibfield{author}{\bibinfo{person}{Po-Han Lin}, \bibinfo{person}{Tzu-Chien Liu}, {and} \bibinfo{person}{Fred Paas}.} \bibinfo{year}{2017}\natexlab{}.
\newblock \showarticletitle{Effects of spell checkers on English as a second language students’ incidental spelling learning: a cognitive load perspective}.
\newblock \bibinfo{journal}{\emph{Reading and Writing}}  \bibinfo{volume}{30} (\bibinfo{year}{2017}), \bibinfo{pages}{1501--1525}.
\newblock


\bibitem[Lin et~al\mbox{.}(2022)]%
        {lin2022teaching}
\bibfield{author}{\bibinfo{person}{Stephanie Lin}, \bibinfo{person}{Jacob Hilton}, {and} \bibinfo{person}{Owain Evans}.} \bibinfo{year}{2022}\natexlab{}.
\newblock \showarticletitle{Teaching models to express their uncertainty in words}.
\newblock \bibinfo{journal}{\emph{arXiv preprint arXiv:2205.14334}} (\bibinfo{year}{2022}).
\newblock


\bibitem[Liu et~al\mbox{.}(2023)]%
        {liu2023prudent}
\bibfield{author}{\bibinfo{person}{Genglin Liu}, \bibinfo{person}{Xingyao Wang}, \bibinfo{person}{Lifan Yuan}, \bibinfo{person}{Yangyi Chen}, {and} \bibinfo{person}{Hao Peng}.} \bibinfo{year}{2023}\natexlab{}.
\newblock \bibinfo{title}{Prudent Silence or Foolish Babble? Examining Large Language Models' Responses to the Unknown}.
\newblock
\newblock
\showeprint[arxiv]{2311.09731}~[cs.CL]


\bibitem[Liu et~al\mbox{.}(2022)]%
        {liu-uist2022}
\bibfield{author}{\bibinfo{person}{Vivian Liu}, \bibinfo{person}{Han Qiao}, {and} \bibinfo{person}{Lydia~B. Chilton}.} \bibinfo{year}{2022}\natexlab{}.
\newblock \showarticletitle{Opal: Multimodal Image Generation for News Illustration}.
\newblock \bibinfo{journal}{\emph{Proceedings of the 35th Annual ACM Symposium on User Interface Software and Technology}} (\bibinfo{year}{2022}).
\newblock


\bibitem[Louie et~al\mbox{.}(2020)]%
        {louie-chi2020}
\bibfield{author}{\bibinfo{person}{Ryan Louie}, \bibinfo{person}{Andy Coenen}, \bibinfo{person}{Cheng-Zhi~Anna Huang}, \bibinfo{person}{Michael Terry}, {and} \bibinfo{person}{Carrie~J. Cai}.} \bibinfo{year}{2020}\natexlab{}.
\newblock \showarticletitle{Novice-AI Music Co-Creation via AI-Steering Tools for Deep Generative Models}.
\newblock \bibinfo{journal}{\emph{Proceedings of the 2020 CHI Conference on Human Factors in Computing Systems}} (\bibinfo{year}{2020}).
\newblock


\bibitem[Lundberg et~al\mbox{.}(2018)]%
        {lundberg2018explainable}
\bibfield{author}{\bibinfo{person}{Scott~M Lundberg}, \bibinfo{person}{Bala Nair}, \bibinfo{person}{Monica~S Vavilala}, \bibinfo{person}{Mayumi Horibe}, \bibinfo{person}{Michael~J Eisses}, \bibinfo{person}{Trevor Adams}, \bibinfo{person}{David~E Liston}, \bibinfo{person}{Daniel King-Wai Low}, \bibinfo{person}{Shu-Fang Newman}, \bibinfo{person}{Jerry Kim}, {et~al\mbox{.}}} \bibinfo{year}{2018}\natexlab{}.
\newblock \showarticletitle{Explainable machine-learning predictions for the prevention of hypoxaemia during surgery}.
\newblock \bibinfo{journal}{\emph{Nature biomedical engineering}} \bibinfo{volume}{2}, \bibinfo{number}{10} (\bibinfo{year}{2018}), \bibinfo{pages}{749--760}.
\newblock


\bibitem[Martindale et~al\mbox{.}(2019)]%
        {martindale2019identifying}
\bibfield{author}{\bibinfo{person}{Marianna Martindale}, \bibinfo{person}{Marine Carpuat}, \bibinfo{person}{Kevin Duh}, {and} \bibinfo{person}{Paul McNamee}.} \bibinfo{year}{2019}\natexlab{}.
\newblock \showarticletitle{Identifying fluently inadequate output in neural and statistical machine translation}. In \bibinfo{booktitle}{\emph{Proceedings of Machine Translation Summit XVII Volume 1: Research Track}}. \bibinfo{pages}{233--243}.
\newblock


\bibitem[Mielke et~al\mbox{.}(2022)]%
        {mielke2022reducing}
\bibfield{author}{\bibinfo{person}{Sabrina~J Mielke}, \bibinfo{person}{Arthur Szlam}, \bibinfo{person}{Emily Dinan}, {and} \bibinfo{person}{Y-Lan Boureau}.} \bibinfo{year}{2022}\natexlab{}.
\newblock \showarticletitle{Reducing conversational agents’ overconfidence through linguistic calibration}.
\newblock \bibinfo{journal}{\emph{Transactions of the Association for Computational Linguistics}}  \bibinfo{volume}{10} (\bibinfo{year}{2022}), \bibinfo{pages}{857--872}.
\newblock


\bibitem[Mozannar et~al\mbox{.}(2022)]%
        {mozannar2022reading}
\bibfield{author}{\bibinfo{person}{Hussein Mozannar}, \bibinfo{person}{Gagan Bansal}, \bibinfo{person}{Adam Fourney}, {and} \bibinfo{person}{Eric Horvitz}.} \bibinfo{year}{2022}\natexlab{}.
\newblock \showarticletitle{Reading Between the Lines: Modeling User Behavior and Costs in AI-Assisted Programming}.
\newblock \bibinfo{journal}{\emph{arXiv preprint arXiv:2210.14306}} (\bibinfo{year}{2022}).
\newblock


\bibitem[Mozannar et~al\mbox{.}(2021)]%
        {mozannar-aaai2021}
\bibfield{author}{\bibinfo{person}{Hussein Mozannar}, \bibinfo{person}{Arvindmani Satyanarayan}, {and} \bibinfo{person}{David~A. Sontag}.} \bibinfo{year}{2021}\natexlab{}.
\newblock \showarticletitle{Teaching Humans When To Defer to a Classifier via Examplars}. In \bibinfo{booktitle}{\emph{AAAI}}.
\newblock


\bibitem[Niculescu-Mizil and Caruana(2005)]%
        {niculescu-icml2005}
\bibfield{author}{\bibinfo{person}{Alexandru Niculescu-Mizil} {and} \bibinfo{person}{Rich Caruana}.} \bibinfo{year}{2005}\natexlab{}.
\newblock \showarticletitle{Predicting good probabilities with supervised learning}. In \bibinfo{booktitle}{\emph{Proceedings of the 22nd international conference on Machine learning}}. \bibinfo{pages}{625--632}.
\newblock


\bibitem[OpenAI(2015)]%
        {playground}
\bibfield{author}{\bibinfo{person}{OpenAI}.} \bibinfo{year}{2015}\natexlab{}.
\newblock
\urldef\tempurl%
\url{https://beta.openai.com/playground}
\showURL{%
\tempurl}


\bibitem[Parasuraman and Manzey(2010)]%
        {parasuraman2010complacency}
\bibfield{author}{\bibinfo{person}{Raja Parasuraman} {and} \bibinfo{person}{Dietrich~H Manzey}.} \bibinfo{year}{2010}\natexlab{}.
\newblock \showarticletitle{Complacency and bias in human use of automation: An attentional integration}.
\newblock \bibinfo{journal}{\emph{Human factors}} \bibinfo{volume}{52}, \bibinfo{number}{3} (\bibinfo{year}{2010}), \bibinfo{pages}{381--410}.
\newblock


\bibitem[Pearce et~al\mbox{.}(2022)]%
        {pearce_2022}
\bibfield{author}{\bibinfo{person}{Hammond Pearce}, \bibinfo{person}{Baleegh Ahmad}, \bibinfo{person}{Benjamin Tan}, \bibinfo{person}{Brendan Dolan-Gavitt}, {and} \bibinfo{person}{Ramesh Karri}.} \bibinfo{year}{2022}\natexlab{}.
\newblock \showarticletitle{Asleep at the Keyboard? Assessing the Security of GitHub Copilot’s Code Contributions}. In \bibinfo{booktitle}{\emph{2022 IEEE Symposium on Security and Privacy (SP)}}. \bibinfo{pages}{754--768}.
\newblock
\urldef\tempurl%
\url{https://doi.org/10.1109/SP46214.2022.9833571}
\showDOI{\tempurl}


\bibitem[Perry et~al\mbox{.}(2022)]%
        {perry2022users}
\bibfield{author}{\bibinfo{person}{Neil Perry}, \bibinfo{person}{Megha Srivastava}, \bibinfo{person}{Deepak Kumar}, {and} \bibinfo{person}{Dan Boneh}.} \bibinfo{year}{2022}\natexlab{}.
\newblock \showarticletitle{Do users write more insecure code with AI assistants?}
\newblock \bibinfo{journal}{\emph{arXiv preprint arXiv:2211.03622}} (\bibinfo{year}{2022}).
\newblock


\bibitem[Pudari and Ernst(2023)]%
        {pudari2023copilot}
\bibfield{author}{\bibinfo{person}{Rohith Pudari} {and} \bibinfo{person}{Neil~A. Ernst}.} \bibinfo{year}{2023}\natexlab{}.
\newblock \bibinfo{title}{From Copilot to Pilot: Towards AI Supported Software Development}.
\newblock
\newblock
\showeprint[arxiv]{2303.04142}~[cs.SE]


\bibitem[Radford et~al\mbox{.}(2019)]%
        {radford2019gpt2}
\bibfield{author}{\bibinfo{person}{Alec Radford}, \bibinfo{person}{Jeffrey Wu}, \bibinfo{person}{Rewon Child}, \bibinfo{person}{David Luan}, \bibinfo{person}{Dario Amodei}, {and} \bibinfo{person}{Ilya Sutskever}.} \bibinfo{year}{2019}\natexlab{}.
\newblock \bibinfo{title}{Language models are unsupervised multitask learners}.
\newblock \bibinfo{howpublished}{OpenAI white paper}.
\newblock


\bibitem[Sarkar et~al\mbox{.}(2022)]%
        {sarkar2022like}
\bibfield{author}{\bibinfo{person}{Advait Sarkar}, \bibinfo{person}{Andrew~D Gordon}, \bibinfo{person}{Carina Negreanu}, \bibinfo{person}{Christian Poelitz}, \bibinfo{person}{Sruti~Srinivasa Ragavan}, {and} \bibinfo{person}{Ben Zorn}.} \bibinfo{year}{2022}\natexlab{}.
\newblock \showarticletitle{What is it like to program with artificial intelligence?}
\newblock \bibinfo{journal}{\emph{arXiv preprint arXiv:2208.06213}} (\bibinfo{year}{2022}).
\newblock


\bibitem[Saunders et~al\mbox{.}(2022)]%
        {saunders2022selfcritiquing}
\bibfield{author}{\bibinfo{person}{William Saunders}, \bibinfo{person}{Catherine Yeh}, \bibinfo{person}{Jeff Wu}, \bibinfo{person}{Steven Bills}, \bibinfo{person}{Long Ouyang}, \bibinfo{person}{Jonathan Ward}, {and} \bibinfo{person}{Jan Leike}.} \bibinfo{year}{2022}\natexlab{}.
\newblock \bibinfo{title}{Self-critiquing models for assisting human evaluators}.
\newblock
\newblock
\showeprint[arxiv]{2206.05802}~[cs.CL]


\bibitem[Shrivastava et~al\mbox{.}(2023)]%
        {shrivastava2023llamas}
\bibfield{author}{\bibinfo{person}{Vaishnavi Shrivastava}, \bibinfo{person}{Percy Liang}, {and} \bibinfo{person}{Ananya Kumar}.} \bibinfo{year}{2023}\natexlab{}.
\newblock \bibinfo{title}{Llamas Know What GPTs Don't Show: Surrogate Models for Confidence Estimation}.
\newblock
\newblock
\showeprint[arxiv]{2311.08877}~[cs.CL]


\bibitem[Singh et~al\mbox{.}(2023)]%
        {singh2023confidencecompetence}
\bibfield{author}{\bibinfo{person}{Aniket~Kumar Singh}, \bibinfo{person}{Suman Devkota}, \bibinfo{person}{Bishal Lamichhane}, \bibinfo{person}{Uttam Dhakal}, {and} \bibinfo{person}{Chandra Dhakal}.} \bibinfo{year}{2023}\natexlab{}.
\newblock \bibinfo{title}{The Confidence-Competence Gap in Large Language Models: A Cognitive Study}.
\newblock
\newblock
\showeprint[arxiv]{2309.16145}~[cs.CL]


\bibitem[Sun et~al\mbox{.}(2022)]%
        {ibm}
\bibfield{author}{\bibinfo{person}{Jiao Sun}, \bibinfo{person}{Q.~Vera Liao}, \bibinfo{person}{Michael Muller}, \bibinfo{person}{Mayank Agarwal}, \bibinfo{person}{Stephanie Houde}, \bibinfo{person}{Kartik Talamadupula}, {and} \bibinfo{person}{Justin~D. Weisz}.} \bibinfo{year}{2022}\natexlab{}.
\newblock \showarticletitle{Investigating Explainability of Generative AI for Code through Scenario-Based Design}. In \bibinfo{booktitle}{\emph{27th International Conference on Intelligent User Interfaces}} (Helsinki, Finland) \emph{(\bibinfo{series}{IUI '22})}. \bibinfo{publisher}{Association for Computing Machinery}, \bibinfo{address}{New York, NY, USA}, \bibinfo{pages}{212–228}.
\newblock
\showISBNx{9781450391443}
\urldef\tempurl%
\url{https://doi.org/10.1145/3490099.3511119}
\showDOI{\tempurl}


\bibitem[Tanneru et~al\mbox{.}(2023)]%
        {tanneru2023quantifying}
\bibfield{author}{\bibinfo{person}{Sree~Harsha Tanneru}, \bibinfo{person}{Chirag Agarwal}, {and} \bibinfo{person}{Himabindu Lakkaraju}.} \bibinfo{year}{2023}\natexlab{}.
\newblock \bibinfo{title}{Quantifying Uncertainty in Natural Language Explanations of Large Language Models}.
\newblock
\newblock
\showeprint[arxiv]{2311.03533}~[cs.CL]


\bibitem[Vaithilingam et~al\mbox{.}(2022)]%
        {vaithilingam-chi2022}
\bibfield{author}{\bibinfo{person}{Priyan Vaithilingam}, \bibinfo{person}{Tianyi Zhang}, {and} \bibinfo{person}{Elena~L Glassman}.} \bibinfo{year}{2022}\natexlab{}.
\newblock \showarticletitle{Expectation vs. Experience: Evaluating the Usability of Code Generation Tools Powered by Large Language Models}. In \bibinfo{booktitle}{\emph{CHI Conference on Human Factors in Computing Systems Extended Abstracts}}. \bibinfo{pages}{1--7}.
\newblock


\bibitem[Van~der Bles et~al\mbox{.}(2019)]%
        {van2019communicating}
\bibfield{author}{\bibinfo{person}{Anne~Marthe Van~der Bles}, \bibinfo{person}{Sander Van Der~Linden}, \bibinfo{person}{Alexandra~LJ Freeman}, \bibinfo{person}{James Mitchell}, \bibinfo{person}{Ana~B Galvao}, \bibinfo{person}{Lisa Zaval}, {and} \bibinfo{person}{David~J Spiegelhalter}.} \bibinfo{year}{2019}\natexlab{}.
\newblock \showarticletitle{Communicating uncertainty about facts, numbers and science}.
\newblock \bibinfo{journal}{\emph{Royal Society open science}} \bibinfo{volume}{6}, \bibinfo{number}{5} (\bibinfo{year}{2019}), \bibinfo{pages}{181870}.
\newblock


\bibitem[Vasconcelos et~al\mbox{.}(2022)]%
        {helena}
\bibfield{author}{\bibinfo{person}{Helena Vasconcelos}, \bibinfo{person}{Matthew Jörke}, \bibinfo{person}{Madeleine Grunde-McLaughlin}, \bibinfo{person}{Tobias Gerstenberg}, \bibinfo{person}{Michael Bernstein}, {and} \bibinfo{person}{Ranjay Krishna}.} \bibinfo{year}{2022}\natexlab{}.
\newblock \bibinfo{title}{Explanations Can Reduce Overreliance on AI Systems During Decision-Making}.
\newblock
\newblock
\urldef\tempurl%
\url{https://doi.org/10.48550/ARXIV.2212.06823}
\showDOI{\tempurl}


\bibitem[Vaswani et~al\mbox{.}(2017)]%
        {vaswani2017attention}
\bibfield{author}{\bibinfo{person}{Ashish Vaswani}, \bibinfo{person}{Noam Shazeer}, \bibinfo{person}{Niki Parmar}, \bibinfo{person}{Jakob Uszkoreit}, \bibinfo{person}{Llion Jones}, \bibinfo{person}{Aidan~N Gomez}, \bibinfo{person}{{\L}ukasz Kaiser}, {and} \bibinfo{person}{Illia Polosukhin}.} \bibinfo{year}{2017}\natexlab{}.
\newblock \showarticletitle{Attention is All you Need}. In \bibinfo{booktitle}{\emph{Advances in Neural Information Processing Systems}}, Vol.~\bibinfo{volume}{30}.
\newblock


\bibitem[Wang et~al\mbox{.}(2021)]%
        {wang-elsevier2021}
\bibfield{author}{\bibinfo{person}{Danding Wang}, \bibinfo{person}{Wencan Zhang}, {and} \bibinfo{person}{Brian~Y Lim}.} \bibinfo{year}{2021}\natexlab{}.
\newblock \showarticletitle{Show or suppress? Managing input uncertainty in machine learning model explanations}.
\newblock \bibinfo{journal}{\emph{Artificial Intelligence}}  \bibinfo{volume}{294} (\bibinfo{year}{2021}), \bibinfo{pages}{103456}.
\newblock


\bibitem[Wickens et~al\mbox{.}(2015)]%
        {wickens2015complacency}
\bibfield{author}{\bibinfo{person}{Christopher~D Wickens}, \bibinfo{person}{Benjamin~A Clegg}, \bibinfo{person}{Alex~Z Vieane}, {and} \bibinfo{person}{Angelia~L Sebok}.} \bibinfo{year}{2015}\natexlab{}.
\newblock \showarticletitle{Complacency and automation bias in the use of imperfect automation}.
\newblock \bibinfo{journal}{\emph{Human factors}} \bibinfo{volume}{57}, \bibinfo{number}{5} (\bibinfo{year}{2015}), \bibinfo{pages}{728--739}.
\newblock


\bibitem[Wu et~al\mbox{.}(2023)]%
        {wu2023autogen}
\bibfield{author}{\bibinfo{person}{Qingyun Wu}, \bibinfo{person}{Gagan Bansal}, \bibinfo{person}{Jieyu Zhang}, \bibinfo{person}{Yiran Wu}, \bibinfo{person}{Shaokun Zhang}, \bibinfo{person}{Erkang Zhu}, \bibinfo{person}{Beibin Li}, \bibinfo{person}{Li Jiang}, \bibinfo{person}{Xiaoyun Zhang}, {and} \bibinfo{person}{Chi Wang}.} \bibinfo{year}{2023}\natexlab{}.
\newblock \bibinfo{title}{AutoGen: Enabling Next-Gen LLM Applications via Multi-Agent Conversation Framework}.
\newblock
\newblock
\showeprint[arxiv]{2308.08155}~[cs.AI]


\bibitem[Yin et~al\mbox{.}(2019)]%
        {yin-chi2019}
\bibfield{author}{\bibinfo{person}{Ming Yin}, \bibinfo{person}{Jennifer~Wortman Vaughan}, {and} \bibinfo{person}{Hanna~M. Wallach}.} \bibinfo{year}{2019}\natexlab{}.
\newblock \showarticletitle{Understanding the Effect of Accuracy on Trust in Machine Learning Models}.
\newblock \bibinfo{journal}{\emph{Proceedings of the 2019 CHI Conference on Human Factors in Computing Systems}} (\bibinfo{year}{2019}).
\newblock


\bibitem[Zhang et~al\mbox{.}(2020)]%
        {zhang-faact2020}
\bibfield{author}{\bibinfo{person}{Yunfeng Zhang}, \bibinfo{person}{Qingzi~Vera Liao}, {and} \bibinfo{person}{Rachel K.~E. Bellamy}.} \bibinfo{year}{2020}\natexlab{}.
\newblock \showarticletitle{Effect of confidence and explanation on accuracy and trust calibration in AI-assisted decision making}.
\newblock \bibinfo{journal}{\emph{Proceedings of the 2020 Conference on Fairness, Accountability, and Transparency}} (\bibinfo{year}{2020}).
\newblock


\bibitem[Zhou et~al\mbox{.}(2023)]%
        {zhou2023navigating}
\bibfield{author}{\bibinfo{person}{Kaitlyn Zhou}, \bibinfo{person}{Dan Jurafsky}, {and} \bibinfo{person}{Tatsunori Hashimoto}.} \bibinfo{year}{2023}\natexlab{}.
\newblock \showarticletitle{Navigating the grey area: Expressions of overconfidence and uncertainty in language models}.
\newblock \bibinfo{journal}{\emph{arXiv preprint arXiv:2302.13439}} (\bibinfo{year}{2023}).
\newblock


\bibitem[Ziegler et~al\mbox{.}(2022)]%
        {Ziegler_2022}
\bibfield{author}{\bibinfo{person}{Albert Ziegler}, \bibinfo{person}{Eirini Kalliamvakou}, \bibinfo{person}{X.~Alice Li}, \bibinfo{person}{Andrew Rice}, \bibinfo{person}{Devon Rifkin}, \bibinfo{person}{Shawn Simister}, \bibinfo{person}{Ganesh Sittampalam}, {and} \bibinfo{person}{Edward Aftandilian}.} \bibinfo{year}{2022}\natexlab{}.
\newblock \showarticletitle{Productivity Assessment of Neural Code Completion}. In \bibinfo{booktitle}{\emph{Proceedings of the 6th ACM SIGPLAN International Symposium on Machine Programming}} (San Diego, CA, USA) \emph{(\bibinfo{series}{MAPS 2022})}. \bibinfo{publisher}{Association for Computing Machinery}, \bibinfo{address}{New York, NY, USA}, \bibinfo{pages}{21–29}.
\newblock
\showISBNx{9781450392730}
\urldef\tempurl%
\url{https://doi.org/10.1145/3520312.3534864}
\showDOI{\tempurl}


\end{thebibliography}

\appendix
\end{document}